\documentclass[12pt]{JHEP3}
\usepackage{nicefrac}
\usepackage[comma,square]{natbib}

\title{Exact chiral ring of AdS$_3/$CFT$_2$}

\preprint{\hepth{0703022}\\TIFR-TH-06-09 \\ YITP-SB-06-09}
\author{Atish Dabholkar\footnote{Email:
atish@theory.tifr.res.in}$^{~1,2}$
and Ari
Pakman\footnote{Email: ari.pakman@stonybrook.edu}$^{~3}$
\\ \\ \\
\it $^1$Department of Theoretical Physics, \\
\it Tata Institute of Fundamental Research,\\
\it Homi Bhabha Rd, Mumbai 400 005, India\\

\it $^2$Laboratoire de Physique Th\'eorique et Hautes Energies (LPTHE)\\
\it{Tour 24-25, 5\` eme \'etage, Boite 126, 4 Place Jussieu, 75252
Paris Cedex 05}\\
\it{Unit\'e Mixte de Recherche (UMR 7589)}\\
\it Universit\'e Pierre et Marie Curie-Paris 6; CNRS; Universit\'e
Denis Diderot-Paris 7\\

\it $^3$C.N. Yang Institute for Theoretical Physics,\\
\it Stony Brook University, \\
\it Stony Brook, NY 11794-3840, USA}

\abstract{
We carry out an exact worldsheet computation of tree level
three-point correlators of chiral operators in type IIB string theory
on $AdS_3 \times S^3 \times T^4$ with  NS-NS flux. We present
a simple representation for the string chiral operators in the
coordinate basis of the dual boundary CFT. Striking cancelations
occur between the three-point functions  of the \h3\  and the $SU(2)$
WZW models which result in a simple factorized form for the final
correlators. We show,  by fixing a single free parameter in the \h3\ WZW model,
that the fusion rules and the  structure constants of the $N=2$ chiral
ring in the  bulk  are in precise agreement with earlier computations
in the boundary CFT of the symmetric product of~$T^4$ at the orbifold point in the large N limit.}

\newcommand{\nn}{\nonumber}

\def\d{\partial}
\newcommand{\rb}[1]{\raisebox{1.5ex}[0pt]{#1}}

\def\a{\alpha}

\def\h{\eta}

\def\up{{\Upsilon}}

\def\CM{{M^4}}

\def\O{{\cal O}}

\def\IC{\relax\hbox{$\inbar\kern-.3em{\rm C}$}}

\def\IC{{\bf C}}

\def\bea{\begin{eqnarray}}
\def\eea{\end{eqnarray}}
\def\be{\begin{equation}}
\def\ee{\end{equation}}
\def\ba{\begin{align}}
\def\ea{\end{align}}
\def\bse{\begin{subequations}}
\def\ese{\end{subequations}}
\def\1F1{{}_1\!F_1}
\def\2F0{{}_2\!F_0}

\def\slr{$SL(2,R)$}
\def\ve{\epsilon}
\def\bi{b^{-1}}

\def\ni{\noindent}
\def\ga{\gamma}
\def\Ga{\Gamma}
\def\bm{\bar{m}}
\def\bx{\bar{x}}
\def\by{\bar{y}}
\def\bz{\bar{z}}

\def\sl{$SL(2,R)$}

\def\nn{\nonumber}

\def\a{\alpha}

\def\h3{$\textrm{H}_3^+$}
\def\d{{\partial}}

\def\IC{{\mathbb C}}

\def\lbldef#1#2{\expandafter\gdef\csname #1\endcsname {#2}}

\def\href#1#2{#2}

\newcommand{\beq}{\begin{equation}}
\newcommand{\eeq}{\end{equation}}
\newcommand{\ber}{\begin{eqnarray}}
\newcommand{\eer}{\end{eqnarray}}

\def\be{\begin{eqnarray}}
\def\ee{\end{eqnarray}}

\keywords{AdS/CFT}

\begin{document}
%\maketitle \setcounter{tocdepth}{2}
%\tableofcontents

\section{Introduction}
Within the family of $AdS_{n+1}/CFT_n$  holographically dual
theories \cite{Maldacena:1997re, Gubser:1998bc, Witten:1998qj,
Aharony:1999ti}, the
$AdS_3/CFT_2$ case has a concrete realization in the form of a duality between
string theory in $AdS_3 \times S^3 \times \CM $, where    $\CM$ is either a torus $T^4$
or a $K3$ surface, and a two-dimensional CFT
in the moduli space of a
non-linear sigma model
whose target space is the symmetric product of $\CM$.

In the heuristic derivation of this $AdS_3/CFT_2$ correspondence
via the near horizon of the $D1/D5$ system in type IIB 10D
supergravity~\cite{Maldacena:1997re} (see \cite{David:2002wn, Martinec} for reviews),
there is RR flux through the
$S^3$ factor of the $AdS_3 \times S^3 \times {\CM}$ geometry,
which makes the theory more difficult to
quantize \cite{Berkovits:1999im,Gotz:2006qp}.
 It is convenient to use an S-dual description~\cite{Maldacena:1998bw},
where the D1/D5 system becomes a system of $Q_1$ fundamental strings and $Q_5$ NS5
branes.
%\footnote{More precisely, the D1/D5 description is valid when
% $l^4/\alpha'^2=g_6^2 Q_1  Q_5 \gg 1$ and
%$g_s = g_6 \sqrt{Q_1/Q_5} \ll 1$, where $g_6= g_s/\sqrt{\tilde{v}}$ and $\sqrt{\tilde{v}}$ is the
%volume of the four-dimensional space $T^4$ or $K3$ in $\alpha'$ units.
%When this condition is
%not met, the S-dual description in terms of NS5/F1 is valid, since
%it maps $g_s \rightarrow 1/g_s$.}
The near horizon geometry is
still $AdS_3 \times S^3 \times \CM$, but the RR flux is traded for
$k=Q_5$ units of NS flux through $AdS_3$ and through~$S^3$, and
the common radius of $AdS_3$ and $S^3$ is  $\sqrt{\alpha' k}$. The
resulting model  has an exact  worldsheet description through
level $k$ supersymmetric \slr\ and $SU(2)$ WZW models.
This allows  a much more detailed treatment of  the bulk theory,
that goes beyond the leading supergravity approximation.
In this controlled setting, the emergence of the two-dimensional superconformal
symmetry of the dual theory from the  worldsheet has been
studied  in~\cite{Giveon:1998ns,  deBoer:1998pp, Kutasov:1999xu, Giveon:2003ku}.

String propagation in this background
is of interest because it
is related to the Strominger-Vafa black hole \cite{Strominger:1996sh}
and one can construct black hole solutions by taking quotients
\cite{Banados:1992wn,David:2002wn}.

Checks of the $AdS_3/CFT_2$ duality in this background
focused so far in comparing the moduli space \cite{Dijkgraaf:1998gf},
and the spectrum of both
theories \cite{Maldacena:1998bw, deBoer:1998ip, Kutasov:1998zh, Argurio:2000tb},
and to our knowledge no successful comparison of dynamical quantities was performed yet,
of the kind done in  \cite{Freedman:1998tz, Lee:1998bx} for the $N=4$ SYM / $AdS_5 \times S^5$ duality.
In this paper we make a step in that direction. We  show that the fusion rules and the structure
constants of the entire $N=2$ chiral ring of the symmetric product, computed at the orbifold point,
agree precisely with computations in the   $AdS_3 \times S^3 \times \CM$ worldsheet.
The correlators for the $AdS_3$ factor are
better studied in its Euclidean version, the~$H_3^+= SL(2,C)/SU(2)$ WZW model,
and to find complete agreement with the boundary we should
fix a free parameter in the \h3\ WZW model to a specific value.
In this work we consider only `unflowed' \slr\ representations \cite{Maldacena:2000hw}.
We also consider only $\CM = T^4$ for simplicity, but the results
can be easily extended to $\CM = K3$.

This precise agreement is quite surprising because our computations in the
bulk are carried out at
a point in the moduli space which does not correspond in the boundary
SCFT to the orbifold point of the symmetric product. One expects
that the  latter
has a boundary B-field turned on as theta angles whereas in  the bulk theory the corresponding field is
switched off since there is no background RR field. To go from one to other would require
for example to turn on RR field in the bulk or twist fields in the boundary.
Our results suggest a non-renormalization of these correlators analogous to what
is found in the context of $AdS_5/CFT_4$ even though we have only
half as much supersymmetry in our case.
Clearly, it would be very interesting to understand this agreement
from general principles.

%Another possibility is that the
%bulk string theory corresponds to an iterated symmetric product as suggested
%in \cite{Argurio:2000tb} and that the dictionary between the bulk and
%the boundary is more subtle. In particular, it is possible that the correlators
%of chiral operators in the  `unflowed' sector that we consider here
%are not affected by deformation  by twist fields in the boundary.

%The comparison we perform  allows us to
%resolve some ambiguities present in the literature regarding the form
%of some of the chiral operators of the orbifold point.
%We also are able to fix a free parameter in the \h3\ WZW
%model from the comparison with the boundary correlators.

In the bulk, we express the chiral spectrum in  the
isospin variables $(x, \bar x)$ and $(y, \bar y)$  of the
$SL(2, R)$ and $SU(2)$ current algebras respectively. This
basis makes for an easy comparison between the boundary and bulk
correlators and is also well suited for computing the tensor
products required in the construction of the chiral operators.
A crucial point of the computation is that all the factors in the three-point function
of the $H_3^+$ WZW model which mix the quantum numbers of the three vertices,
cancel against similar factors from the  $SU(2)$ WZW model.\footnote{This
cancelation was first observed in \cite{Aharony:2003vk}
in the related background $SL(2,R)/U(1) \times SU(2)/U(1)$.
A similar cancelation occurs in the product of the Liouville and minimal
models three-point functions~\cite{Zamolodchikov:2005fy,Kostov:2005av}.}

The outline of the paper is as follows.
In section $\S{\ref{boundary}}$ we review the spectrum, operators and correlators
of the chiral sector of the symmetric product of $T^4$.
In section $\S{\ref{Summary}}$,
we review the relevant aspects of the $SL(2,R)$ and $SU(2)$ WZW models and the chiral spectrum
in the $AdS_3 \times S^3 \times T^4$ worldsheet.
In section $\S{\ref{Bulkcorrelators}}$ we show
how the fusion rules and structure constants of the symmetric product
are obtained from the string worldsheet.
Finally,  in section $\S{\ref{Conclude}}$ we discuss
 possible directions for further explorations.
Appendix  \ref{firstap} contains a detailed derivation
 of the chiral states in the bulk in the $(x,\bx)-(y,\by)$ basis.
In appendix \ref{secondap} we elaborate on the relationship
between the three-point functions of the \h3\ and $SU(2)$ WZW models in order
to better understand the cancelations between them.
In particular, we show how the conformal bootstrap
method used in~\cite{Teschner:1997ft} for \h3\ can also be applied to $SU(2)$.

\vskip 1cm

\section{The Chiral Ring of the Symmetric Product\label{boundary}}
The boundary theory is a $(4, 4)$ SCFT on the moduli space of the
symmetrized product $\textrm{Sym}^N(\CM)$ of $N$ copies of $\CM$. Here $N = Q_1 Q_5$ for $\CM =
T^4 $ and $N = Q_1 Q_5 + 1$ for $\CM = K3$.
For simplicity we consider  $\CM = T^4$ but these considerations
generalize easily to $\CM = K3$.

We work at a point in the moduli space of the theory where we can
think of $Sym^N(T^4)$ as an orbifold $(T^4)^N/ S_N$ where $S_N$ is
the symmetric group action on $N$ objects.
Before orbifolding,
the  $(T^4)^N$ theory  has $4N$ free bosons $\phi^{i}_I$,
that coordinatize the space, with $i=1,2,3,4$ and $I=1,\ldots, N$.
Their superpartners are $4N$
Majorana-Weyl fermions $\xi^{i}_I$.
We will define the complex combinations
\be
X_I^1 &=& \frac{\phi_I^1 +i\phi_I^2}{\sqrt{2}} \qquad \qquad X^2_I = \frac{\phi_I^3 +i\phi_I^4}{\sqrt{2}}
\\
\lambda^1_I &=& \frac{\xi_I^1 +i\xi_I^2}{\sqrt{2}} \qquad \qquad \lambda^2_I = \frac{\xi_I^3 +i\xi_I^4}{\sqrt{2}}
%\lambda^{\dagger 1}_I &=& \frac{\xi^1 -i\xi^2}{\sqrt{2}} \qquad \qquad \lambda^{\dagger 2}_I = \frac{\xi^3 -i\xi^4}{\sqrt{2}}
\ee
The fields are normalized as
\be
X^a_I(z)X^{\dagger b}_J(w) &\sim& - \delta^{ab} \delta_{IJ} \, \log(z-w)\,, \\
\lambda^{a}_I(z) \lambda^{\dagger b}_J(w) &\sim& \frac{\delta^{a b } \delta_{IJ}}{z-w}
\qquad \qquad a,b =1,2\,.
\ee
%There are similar anti-holomorphic fields $\bar{\phi}^i_I$ and $\bar{\lambda}^i_I$.
These fields form a representation of the $N=4$ superconformal algebra with $c=6N$,
with generators
\be
\label{em}
T(z) &=& -\partial X_{I}^{\dagger i} \, \partial X_{I}^{i}
-{1 \over 2} \, \lambda_{I}^{\dagger a} \,\partial \lambda_{I}^{a}
+{1 \over 2} \, \partial \lambda_{I}^{\dagger a} \,\lambda_{I}^{a}
\nn \\
J^1 &=& -\frac{i}{2} \, (\lambda^{1}_I  \lambda^{2}_I +  \lambda^{ \dagger 1}_I \lambda^{ \dagger 2}_I)
\nn \\
J^2 &=& \frac{1}{2} \, (   \lambda^{ \dagger 1}_I \lambda^{ \dagger 2}_I - \lambda^{1}_I  \lambda^{2}_I)
\nn \\
J^3 &=& \frac12 \, (\lambda^{1}_I  \lambda^{\dagger 1}_I  + \lambda^{2}_I \lambda^{\dagger 2}_I )
\\
G^{a} &=& \sqrt{2} \left[ \begin{array}{r} i\lambda^{ 1}_I \\  -\lambda^{\dagger 2}_I \end{array} \right] \partial X^{\dagger 1}_I +
\sqrt{2} \left[ \begin{array}{r} i\lambda^{ 2}_I \\  \lambda^{\dagger 1}_I \end{array} \right] \partial X^{\dagger 2}_I \,\,.
\nn \\
G^{ \bar{a}} &=& \sqrt{2} \left[ \begin{array}{r} i\lambda^{\dagger 1}_I \\  \lambda^{2}_I \end{array} \right] \partial X^1_I +
\sqrt{2} \left[ \begin{array}{r} i\lambda^{\dagger 1}_I \\ -\lambda^{1}_I \end{array} \right] \partial X^2_I
\nn
\ee
There is a similar antiholomorphic copy of all the fields and the algebra.
The  global part of the $SU(2)$ R-symmetry algebra $J^i$
along with the  antiholomorphic $SU(2)$  correspond to the $SO(4) \sim {SU}(2)\times {SU(2)}$
isometries of the $S^3$ factor  in the bulk.

%It is convenient to bosonize the fermions by introducing
%$\Phi^a_I(z, \bar z) = \phi^a_I(z) + \bar\phi^a_I(\bar z)$ so that
%\begin{equation}\label{leftbosonize}
%    \psi_{I}^{+ a}(z)=e^{i\phi_{I}^{a}(z)}, \; \psi_{I}^{-a}(z)=
%\epsilon_{ab} \, e^{-i\phi_{I}^{b}(z)}
%\end{equation}
%for the left-movers and
%\begin{equation}\label{rightbosonize}
%   \bar{\psi}_{I}^{+a}(\bar{z})=e^{i\bar{\phi}_{I}^{a}(\bar{z})},
%\; \psi_{I}^{-a}(\bar{z})=\epsilon_{ab}
%\,e^{-i\bar{\phi}_{I}^{b}(\bar{z})}
%\end{equation}
%for the right-movers. Altogether, now there are $6N$ bosons on the
%left and the right which we assemble into $\Phi^M_I(z, \bar z)$ so
%that they correspond to $X^{a\dot a}_I$ for $M = 1, 2, 3, 4$, and to
%$\Phi^a_I$ for $M=5, 6$.
We will be interested in (c,c) and (a,a) fields under an $N=2$ subalgebra,
satisfying \cite{Lerche:1989uy}
\be
\Delta = Q \qquad \qquad \bar{\Delta} = \bar{Q}
\ee
and
\be
\Delta = -Q \qquad \qquad \bar{\Delta} = -\bar{Q}
\ee
respectively, where $\Delta, \bar{\Delta}$ are the conformal
dimensions and $Q, \bar{Q}$ are the charges under~$J^3, \bar{J}^3$.\footnote{In the standard normalization of the
 $N=2$ subalgebra, the  $U(1)$ R-current is $2J^3$.}

\subsection{Chiral Spectrum \label{Spectrum}}

%$\textrm{Sym}^N(M^4)$

The Hilbert space of the symmetric orbifold is
the direct sum of twisted sectors, each sector corresponding
to a conjugacy class of $S_N$.
The latter can be represented by
disjoint cyclic permutations  of various lengths $n_i$,
\be
(n_1)^{N_1}(n_2)^{N_2} \ldots (n_r)^{N_r}
\ee
such that
\be
\sum _i n_i N_i = N \,.
\ee
Twisted sectors are thus classified by
the various ways of partitioning the integer
$N$ in terms of smaller integers.
The full twist operator
of a given conjugacy class can then be built as $S_N$ invariant
combinations of the $Z_{n_i}$ twist operators that generate  cycles
of lengths $n_i$.

Chiral states in a manifold are in correspondence with the
elements of its Dolbeault cohomology~\cite{Witten:1982df}.
Under this correspondence an element of $H^{p, q}$
corresponds to a chiral operator with chiral charge $p$ on the left
and $q$ on the right.
For a symmetric product of a manifold $\CM$, the cohomology can be expressed
as a Fox space of free particles~\cite{Vafa:1994tf}. In the AdS/CFT context,
this acquires  a physical meaning in terms of the number of particles
in the gravity side~\cite{Maldacena:1998bw,deBoer:1998ip}.
The first quantized spectrum of the symmetric product corresponds to the second quantized
spectrum of the string dual.
Chiral vertex operators representing BPS single particle states
in first-quantized string theory, correspond in the boundary
to chiral primaries in the conjugacy class of a  {\it single} $Z_n$ cycle.

For a $Z_n$ cycle, the generator of the orbifold group acts
cyclically on the fields $\{X_I^a\}$ for $I = 1, \ldots, n$  taking
$X_I^a \to X_{I+1}^a$ for $I <n$ and $X_{n}^a \to X_1^a$,
and similarly on the $\lambda_I^{a}$'s.  To
analyze the twist fields, we can  diagonalize this
cyclic action by
%. We can define a field
%\begin{equation}
%\label{yo}
%    Y(z, \bar z) = \sum_{I=1}^{n} \Phi_I(z, \bar z)
%\end{equation}
%which is real and invariant under this twist. In addition, we can
defining  the  fields
\be
\label{Y}
Y^a_l(z) &=& \frac{1}{\sqrt{n}} \, \sum_{I =1}^n e^{-\frac{2\pi i l I}{n}} X^a_I  \qquad a=1,2 \\
y^{a}_l(z) &=& \frac{1}{\sqrt{n}} \sum_{I =1}^n e^{-\frac{2\pi i l I}{n}} \lambda_I^{a}(z)
\label{yl}
\ee
for $l = 0, 1, 2, \ldots (n-1)$ .
These fields are orthogonal,
\be
Y^a_l(z) Y^{\dagger b}_m(w) &\sim& - \delta^{a b} \delta_{lm} \, \log(z-w) \,,
\\
y^{a}_l(z) y^{\dagger b}_m(w) &\sim& \frac{\delta^{a b} \delta_{lm} }{z-w} \,.
\ee
Therefore we have $2n$ independent complex bosons and fermions
with  boundary conditions
\be
\label{yboundary}
    Y_l^a(ze^{2\pi i}) &=& e^{\frac{2\pi i l}{n}} Y_l^a(z) \\
\label{yboundary2}
 y_l^{a}(ze^{2\pi i}) &=& e^{\frac{2\pi i l}{n}} y^{a}_l(z) \,.
\ee
For each field $Y^a_l$, with $l>1$, this sector is created by the action of a twist field $\sigma^a_l$, with
conformal dimension $\Delta = \frac{l}{2n} (1- \frac{l}{n})$ \cite{Dixon:1986qv}.
For the fermions, we can bosonize them as
\be
y_l^1 = e^{iF_l^1} \qquad \qquad y_l^{\dagger 1} = e^{-iF_l^1}\,, \\
y_l^2 = e^{iF_l^2} \qquad \qquad y_l^{\dagger 2} = e^{-iF_l^2}\,,
\ee
with $F_l^a(z)F_m^b(w) \sim -\delta_{lm}\delta^{ab} \log(z-w)$, and their
twist fields  are $e^{\frac{il}{n}F_l^1}e^{\frac{il}{n}F^2_l}$, with conformal dimension~$\frac{l^2}{n^2}$.
Collecting the factors for all the fields, the $n$-cycle twist operator is
\be
\Sigma_{(12\ldots n)} =
 \prod_{l=1}^{n-1} \sigma_l^1(z,\bz) \sigma_l^2(z,\bz)
e^{\frac{il}{n}(F_l^1+ \bar{F}_l^1 )} e^{\frac{il}{n}(F^2_l+ \bar{F}_l^2)} \,,
\label{sigmatwist}
\ee
where we have included also the anti-holomorphic dependence.
For each $l$ in the above product, the conformal dimension of the twist fields is
\be
\Delta_l = \bar{\Delta}_l = 2 \times \frac{l}{2n} (1- \frac{l}{n}) + 2 \times \frac{l^2}{2 n^2} = \frac{l}{n} \,,
\ee
and therefore the conformal dimension of $\Sigma_{(12\ldots n)}$ is
\be
\Delta= \bar{\Delta} = \sum_{l=1}^{n-1} \Delta_l = \frac{n-1}{2}
\ee
The charge of each factor, measured with
\be
J^3 &=&
%\frac12 (\lambda_I^1 \lambda_I^{\dagger 1} +  \lambda_I^2 \lambda_I^{\dagger 2}) \\
%    &=& \frac12  \left( y_l^1 y_l^{\dagger 1} +  y_l^2 y_l^{\dagger 2}\right)
%+  \frac12 \sum_{I=n+1}^{N} (\lambda_I^1 \lambda_I^{\dagger 1} +  \lambda_I^2 \lambda_I^{\dagger 2})
%\\
 %&=&
\frac{i}{2} \sum_{l=0}^{n-1} \left( \partial F_l^1 + \partial F_l^2 \right)
+  \frac12 \sum_{I=n+1}^{N} (\lambda_I^1 \lambda_I^{\dagger 1} +  \lambda_I^2 \lambda_I^{\dagger 2})
\,,
\ee
is
\be
Q_l = \frac{l}{n} = \Delta_l
\ee
so the total charge is $Q=\Delta= \frac{n-1}{2}$, and also $\bar{Q}=\bar{\Delta}= \frac{n-1}{2}$, and therefore $\Sigma_{(12\ldots n)}$ is chiral-chiral.
More generally,  every chiral operator in $T^4$ will give a chiral operator in every twisted sector \cite{Vafa:1994tf},
and~$\Sigma_{(12\ldots n)}$ corresponds to the identity field in $T^4$.
The other (c,c)  fields in each $T^4$ are, for the scalar sector (no sum on $I$),
\be
\lambda^{a}_{I} \bar{\lambda}^{\bar{a}}_{I} \,, \qquad \qquad a, \bar{a} =1,2
\ee
which have $\Delta = Q = \bar{\Delta} = \bar{Q} = 1/2$ and correspond to the four $(1,1)$ forms in $T^4$, and
\be
\lambda^{1}_{I} \lambda^{2}_{I} \bar{\lambda}^{\bar{1}}_{I} \bar{\lambda}^{\bar{2}}_{I}
\label{tlcp}
\ee
which has $\Delta = Q = \bar{\Delta} = \bar{Q} = 1$ and corresponds  to the $(2,2)$ form in $T^4$. In each case, we should
multiply~$\Sigma_{(12\ldots n)}$ by a combination of the above chiral fields invariant
under the cyclic permutation.
But this last requirement is not restrictive enough. It turns out that there is a twofold
ambiguity in the construction
of the chiral operators.
The first ambiguity is related to the fact that both  \cite{David:2002wn,Martinec:1998st}
\be
%\Sigma_{(12\ldots n)}^{\a, \bar{\a}} =
\Sigma_{(12\ldots n)} \sum_{I} \lambda^{a}_{I} \bar{\lambda}^{\bar{a}}_{I}
\label{sumnotindep}
\ee
and \cite{Jevicki:1998bm}
\be
%\Sigma_{(12\ldots n)}^{\a, \bar{\a}} =
\Sigma_{(12\ldots n)} \left(\sum_{I} \lambda^{a}_{I}\right)
\left(\sum_{I} \bar{\lambda}^{\bar{a}}_{I} \right)
\label{sumindep}
\ee
are such that the fermions are invariant under the cyclic permutation.
The second ambiguity is the range of $I$ in the above
sums. In order to obtain an operator invariant under $\Sigma_{(12\ldots n)}$,
one can sum over  $I=1 \ldots N$~\cite{David:2002wn} or $I=1 \ldots n$~\cite{Jevicki:1998rr,Jevicki:1998bm}
(more generally one could sum over  $I=1 \ldots m$ for $m \geq n$).
Similar ambiguities exist for  operators built from the chiral fields~(\ref{tlcp}).

Clearly each option leads to different correlation functions. If we sum as in~(\ref{sumindep}),
then the fermionic contributions will be  factorized
into holomorphic and antiholomorphic contributions, which will not be the case in (\ref{sumnotindep}).
Regarding the range of the sum, if it runs over $I=1\ldots N$, the  fermions will not only
commute with $\Sigma_{(12\ldots n)}$, but also with the spin field associated with any other cycle.
As a consequence,  correlators will factorize into a trivial part involving the fermions, and a part involving
$\Sigma_{(12\ldots n)}$ which will be universal for a given choice of  $Z_n$ cycles.
On the other hand, for  $I=1\ldots n$, the fermions will generally not commute
with the spin fields corresponding to other cycles, and  correlators will depend on the operators
multiplying each twist field.

The comparison we perform in the next sections with the string theory correlators shows that
the correlators in the gravity dual are factorized into holomorphic and
anti-holomorphic contributions, and they do depend on the type of fermionic
dressing of the twist  fields. Therefore, the boundary
operators we should consider are
\be
\Sigma_{(12\ldots n)}^{(0,0)} &=& \Sigma_{(12\ldots n)} \,,
\label{sigma0}
\\
\Sigma_{(12\ldots n)}^{(a, \bar{a})} &=& \Sigma_{(12\ldots n)} \, y_0^a \bar{y}_0^{\bar{a}}\,, \qquad \qquad a,\bar{a} = 1,2\,,
%\frac{1}{n} \Sigma_{(12\ldots n)} \left(\sum_{I=1}^{n} \lambda^{a}_{I}\right)
%\left(\sum_{I=1}^{n} \bar{\lambda}^{\bar{a}}_{I} \right)
\label{sumindep2}
\\
\Sigma_{(12\ldots n)}^{(2,2)} &=&  \Sigma_{(12\ldots n)} \, y_0^1 y_0^2 \bar{y}_0^{\bar{1}} \bar{y}_0^{\bar{2}} \,,
\label{sumindep3}
\\
\Sigma_{(12\ldots n)}^{(0,2)} &=&  \Sigma_{(12\ldots n)} \,  \bar{y}_0^{\bar{1}} \bar{y}_0^{\bar{2}}\,, \qquad \textrm{etc.,}
\label{sumindep4}
\ee
where $y_0^a$ was defined in (\ref{yl}) and $\bar{y}_0^{\bar{a}}$ is its anti-holomorphic counterpart.
The fact that these  operators have the form (\ref{sumindep}) is natural from the
point of view of the orbifold twisting, since the orbifold action on the fermions is diagonalized in the $y_l$ (\ref{yboundary2}),
and the fields $y_0$  are neutral under the orbifold action.
In each twisted sector therefore
the chiral fields can be
constructed from the twist fields and the  $y_0, \bar{y}_0$ fields
which do not suffer any twisting.

%These are (c,c) operators with  $\Delta=\bar{\Delta}= \frac{n-1}{2}, \frac{n}{2}$ and $\frac{n+1}{2}$ respectively.

Using these operators  in the $Z_n$ sectors, the
chiral primaries in the full $S_N$ orbifold theory can be
constructed by symmetrization.
Summing over all
permutations, we obtain, for $n>1$, the final expressions for the scalar chiral
primaries
\be
  O_{n}^{(0,0)}(z,\bar{z})&=&{{1 } \over
{(N!(N-n)!n)^{1 \over 2}}}\sum_{h\in S_{N}}
\Sigma^{(0,0)}_{h(1..n)h^{-1}}(z,\bar{z})\,,
\\
    O_{n}^{(a,\bar{a})}(z,\bar{z})&=&{{1 } \over {(N!(N-n)!n)^{1 \over 2}}}
\sum_{h\in S_{N}} \Sigma_{h(1..n)h^{-1}}^{(a,\bar{a})} (z,\bar{z})\,,
%\omega^{r}_{a\bar{a}}(z,\bar{z}), \,,
\\
    O_{n}^{(2,2)}(z,\bar{z})&=& \,{{1 }
\over {(N!(N-n)!n)^{1 \over 2}}}\sum_{h\in S_{N}}
\Sigma_{h(1..n)h^{-1}}^{(2,2)}(z,\bar{z})\,,
\label{otwo}
\ee
and similar expressions for non-scalar operators.
The corresponding anti-chiral fields are obtained by conjugation
and the prefactors are fixed by normalizing as \cite{Jevicki:1998bm}
\be
\langle\,O_{n}^{(0,0)\,\dagger}(\infty)\, O_{n}^{(0,0)}(0)\rangle &=&
\langle\,O_{n}^{(2,2)\,\dagger}(\infty)\, O_{n}^{(2,2)}(0)\rangle = 1 \,,
\label{boundary2pf}
\\
\langle\,O_{n}^{(a,\bar{a})\,\dagger}(\infty)\, O_{n}^{(b,\bar{b})}(0)\rangle &=& \delta^{a \, b} \delta^{\bar{a} \, \bar{b}} \,,
\label{boundary2pf2}
\ee
and similarly for non-scalar operators.
For $n=1$, the expressions (\ref{sigma0})-(\ref{sumindep4}) are already normalized.
To compare with the string theory computations, it will be
useful to  express $n$ in terms of $h$ defined as
\be
n &=& 2h -1 \,.
\ee
In terms of the variable $h$, the quantum numbers of the three families of chiral operators can be summarized in the following table:
\begin{center}
\begin{tabular}{|l|c|l|}
\hline
Field        & $\Delta = Q  $ & Range of $\Delta$
\\ \hline \hline
$O_{h}^{(0,0)}$ &  $h-1$  & $0,\frac12 \ldots \frac{N-1}{2}$ \\  \hline
$O_{h}^{(a,a)}$ &  $h-1/2$ & $\frac12,1 \ldots \frac{N}{2}$  \\ \hline
$O_{h}^{(2,2)}$ &  $h$  & $1,\frac32 \ldots \frac{N+1}{2}$ \\ \hline
\end{tabular}
%\vskip 0.4 cm
%\renewcommand{\baselinestretch}{0.6}
%\small
%\parbox{4 in}
%{{\bf Table 1}: Fermionic primaries which are local in the $x$ and $y$ variables.
%$\psi(x)$ is defined in (\ref{psidef}), $\chi(y)$ in~(\ref{chidef}) and $S^{\pm}(x,y)$ in (\ref{spdef}) and (\ref{smdef}).}
\end{center}
%\end{table}
\vskip 0.6cm

\ni
%Fortunately the correlators for the form (\ref{sigma0})-(\ref{sumindep3})
%of the $Z_n$ operators were studied in \cite{Jevicki:1998bm}
%and we will be able to use the results of that work.
The form  (\ref{sigmatwist}) for $\Sigma_{(12\ldots n)}$ was useful to find its quantum numbers,
but it is not useful to compute correlation functions of twist
fields corresponding to cycles which have a partial overlap.

Correlators of twist fields in symmetric products were studied
by different authors along complementary lines .
The works~\cite{Lunin:2000yv,Lunin:2001pw}, using the path integral formalism,
computed correlators of fields $O^{(\a, \bar{\a})}_{n}$ with only $\a, \bar{a}=0,2$,
but considered all the elements of the SU(2) multiplet of the $N=4$ algebra, of which our
$N=2$ chiral and anti-chiral fields correspond to the highest and lowest $J^3_0$ eigenvalue.

On the other hand, the work~\cite{Jevicki:1998bm},
whose results we will mostly use in this paper, applied  the conformal
bootstrap to the results of~\cite{Arutyunov:1997gt,Arutyunov:1997gi} and
computed the structure constants of all the  $N=2$ chiral ring,
i.e.,  all the operators $O^{(\a, \bar{\a})}_{n}$ with
$\a, \bar{\a}=0,a,2$.\footnote{It would be interesting to
study the chiral correlators of the symmetric product with the  topological field theory techniques
used in \cite{Cecotti:1992th} for abelian orbifolds.}

\subsection{Fusion rules and Structure Constants \label{Correlators}}
The chiral ring is defined by its fusion rules and its structure
constants. The scalar sector of the (c,c) ring can be shown to be
closed, and its fusion rules are \cite{Jevicki:1998bm}
\be
(0,0) \times (0,0) &=& (0,0) + (2,2) \nn \\
(0,0) \times (2,2) &=& (2,2)
\label{fusionrules}
\\
(0,0) \times (a,a) &=& (a,a) \nn \\
(a,a) \times (a,a) &=& (2,2) \nn
\ee
where  the length  $n$ of the cycle of each operator should be such that
the chiral charge is conserved.
Therefore there are five non-zero structure constants, given by
\cite{Jevicki:1998bm}\footnote{The expression in \cite{Jevicki:1998bm} for the second correlator
in (\ref{bound3pf}) has an additional factor of 2 which was corrected in \cite{Lunin:2001pw}.}
\begin{eqnarray} \nonumber
  \langle\,O_{n+k-1}^{(0,0)\,\dagger}\,
   O_{k}^{(0,0)}\,O_{n}^{(0,0)}\rangle &=& F(N,n,k)
\left[\frac{(n+k-1)^3 }{ n\,k} \right]^{\nicefrac12}\,,
\\
\nonumber
   \langle\,O_{n+k-3}^{(2,2)\,\dagger}\,
O_{k}^{(0,0)}\,O_{n}^{(0,0)}\rangle &=& F(N,n,k)
     \left[\frac{(N-(n+k)+3)}{(N-(n+k)+2)\,n\,k\,(n+k-3)}\right]^{\nicefrac12} \,,
\\
 \langle\,O_{n+k-1}^{(2,2)\,\dagger}\,
O_{k}^{(0,0)}\,O_{n}^{(2,2)}\rangle &=& F(N,n,k)
\left[\frac{ n^{3} }{k \, (n+k-1)}\right]^{\nicefrac12} \,,
\label{bound3pf}
\\
 \nonumber
   \langle\,O_{n+k-1}^{(a,\bar{a})\,\dagger}\,
O_{k}^{(0,0)}\,O_{n}^{(b,\bar{b})}\rangle &=& F(N,n,k)
\left[\frac{n\,(n+k-1) }{k}\right]^{\nicefrac12}\delta^{a b}\delta^{\bar{a}\bar{b}} \,,
\\
\nonumber
   \langle\,O_{n+k-1}^{(2,2)\,\dagger}\,
O_{k}^{(a,\bar{a})}\,O_{n}^{(b,\bar{b})}\rangle &=& F(N,n,k)
\left[ \frac{n\,k }{(n+k-1)}\right]^{\nicefrac12} \xi^{a b}  \xi^{\bar{a}\bar{b}} \,,
%\omega^{r}\,*\omega^{s}
\end{eqnarray}
where
\be
F(N,n,k) &=& \left[ \frac{(N-n)!\,(N-k)! }{(N-(n+k-1))!\,N!} \right]^{\nicefrac12} \,,
\\
\xi &=&
\left(
\begin{array}{rr}
0 & 1 \\
1 & 0
\end{array}
\right) \,,
\ee
and the operators were set at $z=\bar{z}=0,1,\infty$.
All these correlators assume $n,k > 1$, and
in each case it can be checked that the chiral charge is conserved.

Both the fusion rules and the correlators are easily extended
to the nonscalar sector. From the analysis in \cite{Jevicki:1998bm}
it follows that the process $(0,0)\times (0,0) \rightarrow (2,2)$
can only occur  simultaneously in the holomorphic and anti-holomorphic
sector, but the other four processes in (\ref{fusionrules})
can be combined independently among themselves. Moreover,
for these cases the structure
constants are completely factorized into holomorphic
and anti-holomorphic contributions, and are given by multiplying pairwise
the square roots of the structure constants of (\ref{bound3pf}).
Thus the chiral ring is fully characterized by  $1 + 4 \times 4=17$ structure constants.
It can also be shown that the (c,a) ring has the same fusion rules and structure constants
of the (c,c) ring.

To compare with the bulk tree level computations, we fix the charges
and take $N\to \infty$. We also change the labels $n,k$ and express them
in terms of $h_{1,2}$  as
\be
n &=& 2h_1 -1 \,,
\\
k &=& 2h_2 -1 \,.
\ee
Using
\be
\lim_{N \rightarrow \infty } F(N,n,k) = \left(\frac{1}{N}\right)^{\nicefrac12} \,,
\ee
the tree level chiral structure constants become
\be
\label{boundary3pf1}
\langle O_{h_3}^{(0,0) \,\dagger }  O_{h_2}^{(0,0)} O_{h_1}^{(0,0)} \rangle
&=& \left(\frac{1}{N}\right)^{1/2} \, \left[\frac{(2h_3-1 )^3}{(2h_1-1)(2h_2-1)} \right]^{1/2}
\\
\label{boundary3pf2}
\langle O_{h_3}^{(2,2) \,\dagger}  O_{h_2}^{(0,0)} O_{h_1}^{(0,0)} \rangle
&=&
 \left(\frac{1}{N}\right)^{1/2}
\left[\frac{1}{(2h_1-1)(2h_2-1) (2h_3-1)} \right]^{1/2}
\\
\langle O_{h_3}^{(2,2)\,\dagger}  O_{h_2}^{(0,0)} O_{h_1}^{(2,2)} \rangle
&=&
\left(\frac{1}{N}\right)^{1/2}
\left[\frac{(2h_1-1)^3}{(2h_2-1)(2h_3-1)} \right]^{1/2}
\\
\label{boundary3pf4}
\langle O_{h_3}^{(a,\bar{a}) \, \dagger}  O_{h_2}^{(0,0)} O_{h_1}^{(b,\bar{b})} \rangle
&=&
\left(\frac{1}{N}\right)^{1/2}
\left[\frac{(2h_1-1)(2h_3-1)}{(2h_2-1)} \right]^{1/2} \delta^{ab} \delta^{\bar{a}\bar{b}}
\\
\label{boundary3pf5}
\langle O_{h_3}^{(2,2) \, \dagger}  O_{h_2}^{(a,\bar{a})} O_{h_1}^{(b,\bar{b})} \rangle
&=&
\left(\frac{1}{N}\right)^{1/2}
\left[\frac{(2h_1-1)(2h_2-1)}{(2h_3-1)} \right]^{1/2} \xi^{a\,b}  \xi^{\bar{a}\,\bar{b}} \,,
\ee
where $h_3$ is fixed from the conservation of $U(1)$ R-charge,
and is given by~$h_3=h_1+h_2 -2$ for~(\ref{boundary3pf2}) and by~$h_3=h_1+h_2-1$ for the other  cases.
The fusion rules  and these
simple factorized formulae for the structure constants
can be reproduced by a completely
different worldsheet calculation in the string theory dual as we show below.

\vskip 1cm

\section{The  $AdS_3 \times S^3 \times T^4 $ Worldsheet Theory \label{Summary}}

The supersymmetric $SL(2,R)_k$ model has symmetries
generated by the supercurrents $\psi^A + \theta J^A$, $A=1,2,3$.
Their OPEs are
\be
J^A(z) J^B(w) \sim &&
{{k\over 2} \eta^{AB} \over (z-w)^2} + {i\epsilon^{AB} {}_{C} J^C(w) \over z-w}~,
\label{jjope}
\\
J^A(z) \psi^B(w) \sim && {i\epsilon^{AB}{}_{C} \psi^C(w)
\over z-w}~,
\label{jpsi}
\\
\psi^A(z) \psi^B(w) \sim && { {k\over 2}\eta^{AB} \over z-w}~, \ee
where $\epsilon^{123}=1$ and capital letter indices are raised and
lowered with $\eta^{AB}=\eta_{AB}=(++-)$. Similarly, the
supersymmetric $SU(2)_k$ model has supercurrents $\chi^a + \theta
K^a$, $a=1,2,3$, with OPEs \be K^a(z) K^b(w) \sim && {{k\over 2}
\delta^{ab} \over (z-w)^2} + {i\epsilon^{ab}{}_{c}
 K^c(w) \over z-w}~,
 \label{kkope}
 \\
K^a(z) \chi^b(w) \sim && {i\epsilon^{ab}{}_{c}  \chi^c(w)
\over z-w}~,\\
\chi^a(z) \chi^b(w) \sim && { {k\over 2} \delta^{ab} \over z-w}~,
\ee
and lower case indices are raised and lowered with $\delta^{ab}=\delta_{ab}=(+,+,+)$.
We will often use the linear combinations
\be
J^{\pm} &\equiv& J^1 \pm i J^2 \qquad \qquad \psi^{\pm} \equiv \psi^1 \pm i \psi^2 \,, \\
K^{\pm} &\equiv& K^1 \pm i K^2 \qquad \qquad \!\!\! \chi^{\pm} \equiv \chi^1 \pm i \chi^2 \,.
\ee
As usual in supersymmetric WZW models,
it is convenient to split the $J^A, K^a$  currents into
\be
J^A &=& j^A +   \hat{ \jmath}^A \,,
\label{jsplit}
\\
K^a &=& k^a +\hat{k}^a\,,
\label{ksplit}
\ee
where
\be
\hat{\jmath}^A &=& -\frac{i}{k} \epsilon^{A}{}_{BC} \psi^B \psi^C \,, \\
\hat{k}^a &=& -\frac{i}{k} \epsilon^{a}{}_{bc} \chi^b \chi^c \,. \ee
The currents $j^A$ and $k^a$ generate {\it bosonic}  $SL(2,R)$ and
$SU(2)$  affine algebra at levels $k+2$ and $k-2$, respectively, and
commute with the free fermions $\psi^A,\chi^a$.  The latter in turn
form a pair of supersymmetric \slr\ and $SU(2)$ models at levels -2
and +2, whose bosonic currents are $\hat{\jmath}^A$ and~$\hat{k}^a$.
The spectrum and the  interactions of the original level $k$
supersymmetric WZW models are factorized into the bosonic WZW models
and the free fermions. In terms of the split currents the stress
tensor and supercurrent of the worldsheet theory are
\be
T&=& \frac{1}{k}j^Aj_A - \frac{1}{k} \psi^A \d \psi_A + \frac{1}{k}k^ak_a  -
\frac{1}{k} \chi^a \d \chi_a + T(T^4) \,,
\\
T_F&=& \frac{2}{k} ( \psi^A j_A + \frac{2i}{k} \psi^1\psi^2\psi^3) +
\frac{2}{k}(\chi^A k_A - \frac{2i}{k}\chi^1\chi^2\chi^3) + T_F(T^4)
\,,
\label{supercurrent}
\ee
%where $T(\CM)$ and $T_F(\CM)$ are the
%stress tensor and supercurrent of $\CM= T^4/K3$.
and one can check that the central charge adds up to $c=15$.
Let us see now some specifics of the \slr\ and $SU(2)$ models separately.

\subsection{The $SL(2,R)$ Model: Currents And Observables}
A primary field of spin $h$ in the $SL(2,R)_{k+2}$ WZW model satisfies
\be
j^A(z)
\Phi_{h}(x,\bx;w,\bar{w}) \sim - \frac{D_x^A \Phi_{h}(x,\bx;w,\bar{w})}{z-w} \,,
\label{jphi}
\ee
where the operators $D_x^A$ are \be
D_x^- &=& \d_x \,, \\
D_x^3 &=& x \d_x + h \,, \\
D_x^+ &=& x^2 \d_x + 2hx \,,
\ee
and there is a similar
antiholomorphic copy.
We will sometimes  omit writing explicitly the antiholomorphic dependence of the operators.
The conformal dimension of $\Phi_h(x,\bx;z, \bz)$ is
\be
\Delta_h= \bar{\Delta}_h = -\frac{h(h-1)}{k}\,,
\ee
and it can be expanded in modes as
\be
\Phi_h(x,\bx) = \sum_{m,\bm} \Phi_{h,m,\bm} x^{-h-m} \bx^{-h-\bm}\,,
\label{phiexp}
\ee
but the range of the summation is not always
well defined \cite{Kutasov:1999xu}. Yet, the action of the zero
modes of the currents on $\Phi_{h,m,\bm}$  is well defined and can
be read from (\ref{jphi}) to be
\be
j^3_0 \Phi_{h,m,\bm} &=& m
\Phi_{h,m,\bm} \,\,\,\,,
\\
j^{\pm}_0 \Phi_{h,m,\bm} &=& (m \mp (h-1) ) \Phi_{h,m\pm1,\bm}
\,\,\,\,,
\ee
and similarly for the anti-holomorphic currents.  In
this work we will mostly use the $(x,\bx)$ basis. These variables
are interpreted as the local coordinates of the two-dimensional
conformal field theory living in the boundary of $AdS_3$.
%Also, as we will see in section \ref{spectrum}, % and in appendix \ref{tensoring},
%the  tensor products of bosonic and fermionic representations needed
%to build  chiral operators have a very simple form in the $(x,\bx)$
%basis.

The spectrum of the bosonic $SL(2,R)_{k+2}$  was obtained
in~\cite{Maldacena:2000hw}, and consists of delta-normalizable
{\it continuous}
representations, with $h=\frac12 + i \mathbb{R}$ and $m= \a +
\mathbb{Z} \, (\a \in [0,1))$, and non-normalizable {\it discrete} highest/lowest weight
representations, with~$h \in \mathbb{R}$ obeying
\be
\frac12 < h <
\frac{k+1}{2}\,,
\label{hboundf}
\ee
and $m=h,h+1\ldots$ (lowest
weight) or $m=-h,-h-1 \ldots$ (highest weight). Along with these
{\it unflowed} representations, one should include the states
generated from them by spectral flow~\cite{Maldacena:2000hw},
which will be treated in \cite{Dabholkar:2006nn}.

The bound (\ref{hboundf}) on $h$  is slightly
stricter than the bound $0<h<k/2+1$ needed for the no-ghost theorem
to hold \cite{Balog:1988jb, Petropoulos:1989fc, Hwang:1990aq,  Henningson:1990ua, Evans:1998qu, Pakman:2003cu, Pakman:2003kh, Asano:2003qb}.
The stricter  bound is required for
the normalizability of the primary operators \cite{Giveon:1999px},
and its two ends are consistent with  the spectral flow symmetry,
which relates a highest weight representation with spin $h$ to a
lowest weight representation with spin $k/2 +1 -h$~\cite{Maldacena:2000hw}.

Expressions similar to (\ref{jphi}) hold also for the total currents $J^A$ and the fermionic currents $\hat{\jmath}^A$
of the decomposition~(\ref{jsplit}). We will
use the letters $\hat{h}$, $h$ and $H$ to denote the \slr\ spins
associated to the currents $\hat{\jmath}^A$, $j^A$ and $J^A$.
In particular, $H$ is the
conformal dimension of a field $\Phi_H(x,\bx)$ in
the dual CFT \cite{Giveon:1998ns}.

The OPEs like (\ref{jphi})  between the currents $J^A$ and a field
$\Phi_H(x)$  can be expressed in a compact way by means of the
current
\be J(x;z) = -J^+(z) +2x J^3(z) -x^2 J^-(z)\,
\ee
as
\be
J(x_1;z) \Phi_H(x_2;w) \sim \frac{1}{z-w}\left[ (x_1-x_2)^2 \d_{
x_2}-2H(x_1-x_2)\right] \Phi_h(x_2;w) \,.
\label{jprimary}
\ee
Similarly, the OPEs (\ref{jjope}) between  the currents $J^A$ can be
also expressed through  $J(x,z)$ as \be J(x_1;z) J(x_2;w) \sim k
\frac{(x_1-x_2)^2}{(z-w)^2} + \frac{1}{z-w}\left[ (x_1-x_2)^2 \d_{
x_2}+2(x_1-x_2)\right] J(x_2;w) \,, \ee and from here we see that
$J(z;x)$ is not an \slr\ primary due to the  first term. On the
other hand, its superpartner,
 \be
\psi(x;z) = -\psi^+(z) +2x \psi^3(z) -x^2\psi^-(z)\,, \label{psidef}
\ee
satisfies
\be J(x_1;z) \psi(x_2;w) \sim  \frac{1}{z-w}\left[
(x_1-x_2)^2 \d_{ x_2}+2(x_1-x_2)\right] \psi(x_2;w) \,,
\ee
which
follows from (\ref{jpsi}). Comparing with (\ref{jprimary}), we see
that the  field $\psi(x;z)$ is an $SL(2,R)$  primary with $H=\hat{h}=-1$. It
will appear below in the construction of the chiral operators.

As  in (\ref{jsplit}), the current $J(x;z)$ can be
split into purely bosonic and fermionic terms as
\be J(x;z) = j(x;z)
+ \hat{\jmath}(x;z) \,,
\label{jxsplit}
\ee
where
\be
\hat{\jmath}(x;z) = -\hat{\jmath}^+(z) +2x \hat{\jmath}^3(z) -x^2
\hat{\jmath}^-(z)\,,
\ee
and  $j(x;z)$ is similarly expressed in
terms of $j^A$.

\vskip 1cm
\subsection{The $SU(2)$ Model: Currents And Observables}

The bosonic $SU(2)_{k-2}$ WZW model has primaries $V_{j,m,\bm}$ with
$m,\bm=-j,\ldots,+j$, and the spin $j$ is bounded by \cite{Zamolodchikov:1986bd,Gepner:1986wi}
\be
0 \leq j \leq \frac{k-2}{2} \,.
\label{jboundf}
\ee
The conformal dimension of $V_{j,m,\bm}$ is
\be
\Delta = \bar{\Delta} = \frac{j(j+1)}{k}\,.
\ee
Similarly to the $x,\bx$ variables of the
\slr\ model, isospin coordinates $y,\by$ can be introduced for
$SU(2)$ \cite{Zamolodchikov:1986bd},  and  the primaries are summed
into
\be
V_j(y,\by) = \sum_{m=-j}^{j} V_{j,m,\bm} y^{-m+j} \by^{-\bm
+j} \,\,\,.
\label{yexpansion}
\ee
The action of the $k^a$ currents on $V_j(y;z)$ is
\be
k^a(z) V_{j}(y;w) \sim - \frac{P_y^a V_j(y;w)}{z-w}  \,,
\label{kv}
\ee
where the differential operators
\be
P_y^- &=& -\d_y \\
P_y^3 &=& y \d_y - j \\
P_y^+ &=& y^2 \d_y - 2jy
\ee
are the $SU(2)$ counterparts of
$D_{x}^A$, and there is a  similar antiholomorphic copy. The  action
of the zero modes of $k^a$  on $V_{j,m,\bm}$ can be read from
(\ref{kv}) to be
\be k^3_0 V_{j,m,\bm} &=& m V_{j,m,\bm}
\label{k3action}
\\
k^{\pm}_0 V_{j,m,\bm} &=& (\pm m +1 +j ) V_{j,m\pm1,\bm} \,\,\,\,\, (m\neq \pm j)
\label{kpmaction}
\\
k^{+}_0V_{j,j,\bm} &=&
k^{-}_0V_{j,-j,\bm} = 0
 \ee
and similarly for $\bar{k}^a_0$.
There are similar expressions for $\hat{k}^a$ and $K^a$, and
we will denote by $\hat{\jmath}$, $j$ and $J$ the
spins associated to $\hat{k}^a$, $k^a$ and $K^a$.
Defining now the current
\be K(y;z)
= -K^+(z) +2y K^3(z) +y^2 K^-(z) \,,
\ee the OPEs (\ref{kkope})
and
 the $K^a$ version of (\ref{kv}) can be expressed as
 \be K(y_1;z) K(y_2;w) &\sim&  -k
\frac{(y_1-y_2)^2}{(z-w)^2} + \frac{1}{z-w} \left[ (y_1-y_2)^2 \d_{
y_2}+2(y_1-y_2)\right] K(y_2;w)  \,,
\\
K(y_1;z) V_J(y_2;w) &\sim& \frac{1}{z-w}\left[ (y_1-y_2)^2 \d_{ y_2}+2J(y_1-y_2)\right] V_J(y_2;w) \,.
\label{kprimary}
\ee
The superpartner of $K(y)$,
\be
\chi(y;z) = -\chi^+(z) +2y \chi^3(z) +y^2\chi^-(z) \,,
\label{chidef}
\ee
is an $SU(2)$ primary field of spin $J=\hat{\jmath}=1$, which satisfies
\be
K(y_1;z) \chi(y_2;w) \sim \frac{1}{z-w}\left[ (y_1-y_2)^2 \d_{ y_2}+2(y_1-y_2)\right] \chi(y_2;w)
\ee
and will appear in the chiral operators below.

Finally, the current $K(y)$ can be split as
\be
K(y;z) = k(y;z) + \hat{k}(y;z) \,,
\label{kxsplit}
\ee
where
\be
\hat{k}(y;z) = -\hat{k}^+(z) +2y \hat{k}^3(z) +y^2 \hat{k}^-(z)
\ee
and $k(y;z)$ is similarly expressed in terms of  $k^a$.

\subsection{Ramond Sector \label{Ramond}}

It is convenient to consider the
Ramond sector of the \slr\ and $SU(2)$ models together.
For this, let us bosonize the $\psi^A, \chi^a$ fermions as
\be
\d H_1 &=& \frac{2}{k} \psi^2 \psi^1 \,, \\
\d H_2 &=& \frac{2}{k} \chi^2 \chi^1 \,, \\
\d H_3 &=& \frac{2}{k} i \psi^3 \chi^3 \,.
\ee
We normalize the  four fermions of $T^4$, $\eta^i, \, i=1\ldots4$, as
\be
\eta^i(z)\eta^j(w) \sim \frac{\delta^{ij}}{z-w}\,,
\ee
and they can be bosonized  as
\be
\d H_4 &=& \eta^2 \eta^1 \,,
\\
\d H_5 &=& \eta^4 \eta^3 \,.
\ee
All bosons are normalized as
\be
H_{i}(z)H_{j}(w) \sim -  \delta_{ij} \log(z-w) \,.
\ee
In order to get the correct anticommutation
among the fermions in their bosonized form, we should
also introduce  proper cocycles  \cite{Kostelecky:1986xg}.
For that, we first define the number operators
\be
N_{i} = i \oint \d H_{i} \,,
\ee
and then work in terms of bosons redefined as
%\be
%\hat{h}_1 &=& H_1 \\
%\hat{H}_2 &=& H_2 + \pi N_1 \\
%\hat{H}_3 &=& H_3 + \pi N_1 + \pi N_2 \\
%\hat{H}_4 &=& H_4 + \pi N_1 + \pi N_2 + \pi N_3\\
%\hat{H}_5 &=& H_5 + \pi N_1 + \pi N_2  + \pi N_3+ \pi N_4
%\ee
\be
\hat{H}_i = H_i + \pi \sum_{j<i} N_j \,.
\ee
The fermions are expressed in terms of $\hat{H}_i$ as
\be
e^{\pm i \hat{H}_1} &=& \frac{\psi^1 \pm i \psi^2}{\sqrt{k}} \qquad
e^{\pm i \hat{H}_2} = \frac{\chi^1 \pm i \chi^2}{\sqrt{k}} \qquad
e^{\pm i \hat{H}_3} =  \frac{\chi^3 \mp  \psi^3}{\sqrt{k}} \,\,, \\
e^{\pm i \hat{H}_4} &=& \frac{\eta^1 \pm i \eta^2}{\sqrt{2}} \qquad
e^{\pm i \hat{H}_5} = \frac{\eta^3 \pm i \eta^4}{\sqrt{2}} \,\,\,,
\ee
and the cocycles pick the right signs  using the relation
\be
e^{iaN_{j}} e^{ibH_{j}} =  e^{ibH_{j}}  e^{iaN_{j}} e^{iab} \qquad
\qquad j=1\ldots 4\,\,.
\ee
The Ramond ground state  is created by
acting on the vacuum with the spin fields
\be S(z)= e^{\frac{i}{2}
\sum_{I} \ve_I \hat{H}_I} \,,
\ee
where $\ve_I = \pm 1$, and the GSO
projection imposes the mutual locality condition \be \prod_{I=1}^{5}
\ve_I = +1\,. \label{gso} \ee In particular, the spin fields are
used to build the spacetime supercharges as \be Q = \oint dz
e^{-\frac{\phi}{2}}S(z) \,. \ee BRST invariance  imposes on the
supercharges a constraint which is not present in flat space~\cite{Giveon:1998ns}.
Commutation of $Q$ with the BRST charge
requires that no $(z-w)^{-3/2}$ singularities appear in the OPE
between the supercurrent $T_F$ and $S$. Let us express  $T_F$ in
(\ref{supercurrent}) as
\be T_F = T_F^{\a} + T_F^{\beta} + T_F(T^4)
\,, \label{tfdec}
\ee
where \be T_F^{\a} &=& \frac{1}{\sqrt{k}}
\left[ e^{+i\hat{H}_1} j^{- } + e^{-i\hat{H}_1} j^{+} +  \left(
e^{+i\hat{H}_3 }-e^{-i\hat{H}_3} \right) j^3               \right.
\label{tfa}
\\
 && \qquad \qquad \qquad  \left.
+ \,\, e^{+i\hat{H}_2} k^{- } + e^{-i\hat{H}_2} k^{+} + \left(
e^{+i\hat{H}_3 }+e^{-i\hat{H}_3} \right) k^3   \right] \nn \ee and
\be T_F^{\beta} &=& \frac{1}{\sqrt{k}} \left[   (i\d \hat{H}_2 -i\d
\hat{H}_1 ) e^{-i\hat{H}_3 } +(i\d \hat{H}_2 +i\d \hat{H}_1 )
e^{+i\hat{H}_3 }       \right] \,. \label{tfb}
\ee
Using these expressions, it is easy to check that to avoid
$(z-w)^{-3/2}$ singularities in the  OPE  between   $T_F^{\beta}$
and  $S$,  we must  impose the constraint \be \prod_{I=1}^{3} \ve_I
= +1\,. \label{tsp} \ee Eqs. (\ref{gso}) and (\ref{tsp}) imply that
$\ve_4\ve_5=+1$ in $S(z)$, and leave a total of $8$ supercharges,
which correspond to the 8 supercharges of the global $N=4$
superconformal algebra in the boundary theory~\cite{Giveon:1998ns}.

The currents are expressed in terms of the~$\hat{H}_i$ bosons as
\be
\hat{\jmath}^3 &=& i \d \hat{H}_1 \,, \\
\hat{\jmath}^{\pm} &=& \pm e^{\pm i \hat{H}_1} \left(  e^{- i \hat{H}_3} - e^{+ i \hat{H}_3}
\right) \,, \\
\hat{k}^3 &=&  i \d \hat{H}_2 \,, \\
\hat{k}^{\pm} &=& \mp e^{\pm i \hat{H}_2} \left(  e^{- i \hat{H}_3} + e^{+ i \hat{H}_3}  \right) \,.
\ee
The spin fields provide two $({\bf \frac12}, {\bf \frac12})$ representations
of the $\hat{\jmath}^A,\hat{k}^a$ currents,
with opposite six-dimensional chirality.
Defining
\be
S_{[\ve_{1},\ve_{2},\ve_3]} = e^{ i\frac{\ve_1}{2}\hat{H}_1+  i\frac{\ve_2}{2}\hat{H}_2 +
i\frac{\ve_3}{2}\hat{H}_3 }
\ee
then the $({\bf \frac12}, {\bf \frac12})$ representation with $\ve_1\ve_2\ve_3=+1$ is given by
\be
|\ve_1,\ve_2 \rangle_{+} = (-1)^{\frac{1-\ve_1}{2}} (i)^{\frac{1-\ve_2}{2}} S_{[\ve_1,\ve_2,\ve_1 \ve_2]} |0 \rangle
\ee
and that with $\ve_1\ve_2\ve_3=-1$ is
\be
|\ve_1,\ve_2 \rangle_{-} =  (i)^{\frac{1-\ve_2}{2}} S_{[\ve_1,\ve_2,-\ve_1 \ve_2]} |0 \rangle \,.
\ee
In both cases the zero modes of the currents act as
\be
\hat{\jmath}^3_0 &=&
\left(
\begin{array}{rr}
\frac12 & 0 \\
0 & -\frac12
\end{array}
\right)
\qquad
\hat{\jmath}^+_0 =
\left(
\begin{array}{rr}
0 & 1 \\
0 & 0
\end{array}
\right)
\qquad
\hat{\jmath}^-_0 =
\left(
\begin{array}{rr}
0 & 0 \\
-1 & 0
\end{array}
\right) \,,
\\
\hat{k}^3_0 &=&
\left(
\begin{array}{rr}
\frac12 & 0 \\
0 & -\frac12
\end{array}
\right)
\qquad
\hat{k}^+_0 =
\left(
\begin{array}{rr}
0 & 1 \\
0 & 0
\end{array}
\right)
\qquad
\hat{k}^-_0 =
\left(
\begin{array}{rr}
0 & 0 \\
1 & 0
\end{array}
\right)\,, \ee and the phases that come from the cocycles in
$S_{[\ve_1,\ve_2,\ve_3]}$ are crucial to obtain these  results.
Given a~$({\bf \frac12}, {\bf \frac12})$ representation, the  linear
combination
\be
S(x,y) |0 \rangle = xy |-- \rangle + x|-+\rangle + y|+-\rangle
+ |++\rangle \,
\label{sdef}
\ee
is a well defined primary in the $(x,y)$ basis
for \slr\ and $SU(2)$,
with $H=\hat{h}=-1/2$ and $J=\hat{\jmath}=1/2$.
% The zero modes $\hat{\jmath}^A_0$ and
%$\hat{k}^a_0$  act as  $-D_x^A$ and $-P_y^a$ (see (\ref{jphi}) and (\ref{kv}))
For each chirality, the explicit expressions  of $S(x,y)$ are
\be S^{+}
(x,y) &=& -xyiS_{[--+]} -x S_{[-+-]} +yiS_{[+--]} + S_{[+++]} \,,
\label{spdef}
\\
S^{-}(x,y) &=& +xyiS_{[---]} +xS_{[-++]} + yiS_{[+-+]}+S_{[++-]} \,.
\label{smdef}
\ee

\vskip 1 cm
\ni
In the table below we summarize the properties
of the fields $\psi(x), \chi(y)$ and $S^{\pm}(x,y)$,  defined in
(\ref{psidef}), (\ref{chidef}) and (\ref{spdef}) and (\ref{smdef}).
They all belong to the Hilbert space of the free fermions and will
play an important role below in the construction of the chiral
operators.

%\begin{table}
\begin{center}
\begin{tabular}{|l|c|c|c|l|}
\hline
Field        & $\hat{h}$ & $\hat{\jmath}$  & Sector & Expansion
\\ \hline \hline
$\psi(x) $%= -\psi^+ +2x \psi^3 -x^2\psi^-$
&  -1 &  -  & NS & $-\psi^+ +2x \psi^3 -x^2\psi^-$ \\ \hline
$\chi(y) $
& - & 1 & NS & $-\chi^+ +2y \chi^3 +y^2\chi^- $
\\ \hline
$S^{\pm}(x,y) $ & -1/2 & 1/2 & R & $\mp xyiS_{[--\pm]} \mp xS_{[-+\mp]} + yiS_{[+-\mp]}+S_{[++\pm]}$
\\
\hline
\end{tabular}
%\vskip 0.4 cm
%\renewcommand{\baselinestretch}{0.6}
%\small
%\parbox{4 in}
%{{\bf Table 1}: Fermionic primaries which are local in the $x$ and $y$ variables.
%$\psi(x)$ is defined in (\ref{psidef}), $\chi(y)$ in~(\ref{chidef}) and $S^{\pm}(x,y)$ in (\ref{spdef}) and (\ref{smdef}).}
\end{center}
%\end{table}

\vskip 1cm

\subsection{Spectrum of Chiral Operators \label{Bulkspectrum}}
\label{spectrum}
Chiral operators belong to $SU(2)$ multiplets which satisfy
\be H=J\,.
\ee
A chiral (anti-chiral) operator corresponds to the state with $K^3_0$
eigenvalue $M=J$ ($M=-J$), but it will be convenient
to keep the whole $SU(2)$ multiplet to compute the correlators.
The spectrum of  physical chiral
operators in the worldsheet of the bulk theory was
obtained in \cite{Kutasov:1998zh} in the $m,\bm$ basis.
In appendix A, we rederive it in the $x,\bx$ basis, which is more
appropriate for the computation of correlation functions.

The result is that all the chiral operators are built
from the basic $k-1$ operators
\be
{\cal O}_h(x,y) \equiv  \Phi_{h} (x) V_{h-1}(y) \qquad \qquad \qquad h = 1,\frac32, \ldots \frac{k}{2} \,,
%\label{odef}
\ee
where
$\Phi_{h} (x)$ and $V_{h-1}(y)$
are primaries of the bosonic $SL(2,R)_{k+2}$
and $SU(2)_{k-2}$ models.
Note that $\Delta (\O_h(x,y))=0$ and that the operators cover the whole range of $k-1$ values for $j=h-1$
allowed by the $SU(2)$ bound (\ref{jboundf}).

In the holomorphic sector, there are three families of chiral operators.
In the $-1$ ($-1/2$) picture of the NS (R) sector,
they are obtained  by multiplying  $\O_h(x,y)$   by any
of the operators $ e^{-\phi} \psi(x)$, $e^{-\phi} \chi(y)$ or  $e^{-\frac{\phi}{2}} s^{a}_{-}(x,y)$ ($a=1,2$),
where
\be
s^{1}_{\pm}(x,y) &=& S^{\pm}(x,y) e^{+ \frac{i}{2}(\hat{H}_4- \hat{H}_5)} \,, \\
s^{2}_{\pm}(x,y) &=& S^{\pm}(x,y) e^{- \frac{i}{2}(\hat{H}_4- \hat{H}_5)} \,,
\ee
and $\phi$ comes from the bosonization of the $\beta-\gamma$ ghosts \cite{Friedan:1985ge}.
We will use a tilde ($\, \tilde{}\, $) to denote the representation  of the
operators in the $0$ and $-3/2$ pictures.
We summarize the holomorphic spectrum in the following
table:\footnote{The correspondence
with the notation in \cite{Kutasov:1998zh} is ${\cal W}_{h-1}^- \leftrightarrow \O_h^{0}$,
${\cal X}_{h-1}^+ \leftrightarrow \O_h^{2}$, ${\cal Y}_{h-1}^{\pm} \leftrightarrow \O_h^{a}$.}
\vskip 0.4cm
\begin{center}
\renewcommand{\baselinestretch}{5.3}
\begin{tabular}{|l|r|l|c|c|l|}
\hline
Op.  & Pic. & Expansion & $H=J$ & NSR  & {\small Range of $H$}
\\
\hline \hline
 $\O_h^{0}$  & -1 & $e^{-\phi} \O_h(x,y) \psi(x) $  & & &
\\
\cline{1-3}
 $\tilde{\O}_h^{0}$ & 0 & $\left( (1- h)\hat{\jmath}(x)  + j(x)
+ \frac{2}{k} \psi(x)  \chi_a P^a_y \right) \O_h(x,y)$ & \rb{$h-1=j$}  &  \rb{NS}  & \rb{$0,\frac12 \ldots \frac{k-2}{2}$}
\\
\hline \hline
$\O_h^{a}$  & $-\frac12$ & $e^{-\frac{\phi}{2}} \O_h(x,y) s^a_-(x,y) $ & & &
\\
\cline{1-3}
$\tilde{\O}_h^{a}$  & $-\frac32$ & $- \sqrt{k}(2h-1)^{-1} e^{-\frac{3\phi}{2}} \O_h(x,y) s^a_+(x,y) $
&  \rb{$h-\frac12= j+\frac12$} & \rb{R}  & \rb{$\frac12,1 \ldots \frac{k-1}{2}$}
\\
\hline \hline
 $\O_h^{2}$  & -1 &  $e^{- \phi } \O_h(x,y) \chi(y) $ & & &
\\
\cline{1-3}
$\tilde{\O}_h^{2}$  & 0 & $\left( h \hat{k}(y)   + k(y)+ \frac{2}{k} \chi(y)  \psi_A D^A_x \right) \O_h(x,y)$ &  \rb{$h= j+1$}
& \rb{NS} & \rb{$1, \frac32 \ldots \frac{k}{2}$}
\\
\hline
\end{tabular}
\vskip 0.4 cm
\end{center}
The anti-holomorphic part of
the operators is fixed by multiplying by an anti-holomorphic
 field
$e^{-\bar{\phi}}\bar{\psi}(\bx)$, $e^{-\bar{\phi}} \bar{\chi}(\by)$ or  $e^{-\frac{\bar{\phi}}{2}} \bar{s}^{\bar{a}}_-(\bx,\by)$.
The full chiral operators
have then the form \be
\O_h^{(0,2)} = e^{-\phi -\bar{\phi} } \O_h(x,\bx,y,\by) \psi(x)
\bar{\chi}(\by)
\ee
and so on, giving a total of nine families, whose spectrum and degeneracies can be
compared~\cite{Kutasov:1998zh} with the KK modes of supergravity computed  in \cite{Larsen:1998xm,deBoer:1998ip}.
The scalar sector is composed by $\O_h^{(0,0)}, \O_h^{(a,\bar{a})} $ and $\O_h^{(2,2)}$.

As was studied in \cite{Rastelli:2005ph}, the chiral
operators are also chiral with respect to the $N=2$ superconformal
symmetry of the worldsheet.

The labeling of the  operators  makes explicit the bulk-boundary dictionary
we propose in this work. This dictionary is based on matching the lowest conformal dimension in each family,
and in the degeneracy in the indices~$(a,\bar{a})$,  which correspond  both in the
bulk and the boundary to the elements of $H^{(1,1)}(T^4)$.

Special mention deserves the $h=1$ operator in the
$\O_h^{(0,0)}$ family, which does not seem to have a counterpart in the boundary.
In the zero picture it is
\be
\tilde{\O}_{h=1}^{(0,0)} = j(x)\bar{j}(\bx) \Phi_{h=1}(x,\bx)\,.
\ee
It has conformal dimension zero in the boundary,
and appears in the central extension of the boundary symmetries built from
the string worldsheet \cite{Kutasov:1999xu}. But it fails
to behave as the identity in a correlator, since its insertion in an $n$-point
function does not give the $(n-1)$-point function, as we will see below for $n=3$.
Several properties of this operator were studied in \cite{Giveon:2001up}.

Assuming $Q_5=k$,  the number of operators in each family
in the bulk is $Q_5-1$, less than the $N=Q_1Q_5$  operators in the boundary.
Even though this will not prevent us from performing a successful
comparison of the correlators for those operators present both in the bulk and in the boundary,
a few words on this point are in order.

A complete treatment of the $SL(2,R)$ WZW model
must include the spectrally flowed representations of $SL(2,R)$~\cite{Maldacena:2000hw},
to be considered in \cite{Dabholkar:2006nn}.\footnote{These representations
are necessary in order to obtain a modular invariant
partition function~\cite{Maldacena:2000kv, Israel:2003ry}
(see also~\cite{Henningson:1991jc, Petropoulos:1999nc}), to ensure
that the spacetime energy does not have an unphysical upper bound,
and to properly account for the states corresponding to
the long string excitations \cite{Maldacena:1998uz, Seiberg:1999xz}.}
But including them leads to an infinite number of chiral operators in the bulk.
A resolution to this problem was proposed in~\cite{Hikida:2000ry,Argurio:2000tb}. The idea
is that the spectral flow parameter $w$, which in perturbation theory spans all the integers,  should be restricted to
$0 \leq w \leq Q_1-1$. This 'stringy exclusion principle' \cite{Maldacena:1998bw} is not seen in
the worldsheet because the six dimensional
string coupling is \cite{Giveon:1998ns}
\be
g_6^2 = \frac{Q_5}{Q_1} \,,
\ee
so string perturbation theory needs $Q_1 \gg 1$.
With this prescription there are $Q_1 (k-1)$ operators in the
bulk and the missing $Q_1$ operators can  be explained away along the
lines of~\cite{Seiberg:1999xz}.\footnote{
%If we identify $k=Q_5$ instead, then
%there will be one operator missing per family in each
%of the $Q_1$ spectrally flowed sectors. Such a mismatch could perhaps be
%explained along the lines of \cite{Seiberg:1999xz}.
It was argued in \cite{Son:2003zv} that in the plane wave
limit that the missing chiral operators appear in the continuous representations of $SL(2,R)$.
Notice that if we would identify $Q_5=k-1$ there would be no operators missing. This shift
is allowed  in the large $Q_5$ limit needed for supergravity to be valid.
%Our computation will be independent of this identification since this ambiguity
%will appear only in a single normalization constant.
}
%Another possibility might be to identify $k=Q_5+1$

Given this split between $Q_1$ and $Q_5$ for the  quantum numbers in the string side,
it was further suggested in \cite{Argurio:2000tb} that the boundary theory
is more naturally identified as a deformation of the  iterated symmetric product
\be
\textrm{Sym}^{Q_1}\left( \textrm{Sym}^{Q_5}(\CM)\right)\,.
\label{iterated}
\ee
%which  has the same chiral spectrum as $\textrm{Sym}^{Q_1 Q_5}(\CM)$.
In this setting, the computations of this paper for the $w=0$ sector in
the bulk would correspond to the identity sector in  $\textrm{Sym}^{Q_1}$.
%and the number $N$ in the correlators of section $\S{\ref{boundary}}$ should be identified with $k=Q_5$.

\section{Three-point Correlation Functions\label{Bulkcorrelators}}

\subsection{The Basic Cancelation \label{Cancel}}
Since the basic building block of all the chiral operators is the field $\O_h(x,y)=\Phi_{h} (x) V_{h-1}(y)$,
any three-point correlator among them will involve the value of
\be
\langle \O_{h_1}(x_1 ,y_1)  \O_{h_2}(x_2 ,y_2)   \O_{h_3}(x_3 ,y_3) \rangle \,,
\label{3pfo}
\ee
which  is the product of
\be
\langle \Phi_{h_1}(x_1,\bar{x}_1)
 \Phi_{h_2}(x_2,\bar{x}_2)
 \Phi_{h_3}(x_3,\bar{x}_3)
\rangle = \frac{C_{H}(h_1,h_2,h_3)}{|x_{12}|^{2h_1 + 2h_2 -2h_3}
|x_{23}|^{2 h_2 + 2h_3 -2h_1}
|x_{31}|^{2 h_3 + 2h_1 -2h_2}
}
\label{3pfphi}
\ee
and
\be
\langle V_{j_1}(y_1,\bar{y}_1)
 V_{j_2}(y_2,\bar{y}_2)
 V_{j_3}(y_3,\bar{y}_3)
\rangle &=& C_{S}(j_1,j_2,j_3)  \nn \\
&\times& |y_{12}|^{2j_1 + 2j_2 -2j_3} |y_{23}|^{2 j_2 + 2j_3 -2j_1}
|y_{31}|^{2 j_3 + 2j_1 -2j_2} \,
\label{3pfv}
\ee
evaluated at
\be
j_i= h_i - 1 \qquad \qquad (i=1,2,3)\,.
\label{hjo}
\ee
Eq.(\ref{3pfo}) has no dependence on the $z_i$'s because $\Delta(\O_h) =0$,
and for (\ref{3pfphi})
and~(\ref{3pfv})
the dependence on the $z_i's$ is standard and we have omitted it.

The expressions $C_{H}(h_1,h_2,h_3)$ and $C_{S}(j_1,j_2,j_3)$ are the
three-point functions of the \h3\ and $SU(2)$ WZW models at levels $k+2$ and $k-2$, respectively.
For the $SU(2)$ case, they are~\cite{Zamolodchikov:1986bd}
\be
C_S(j_1,j_2,j_3)=  N_{j_1,j_2,j_3} \sqrt{\ga(b^2)} P(j+1)
\prod_{i=1}^{3}
\frac{P(j-2j_i)}{ P(2j_i) \sqrt{\ga((2j_i+1)b^2)} }\,,
\label{zf3pf}
\ee
where
\be
j &=& j_1+j_2+j_3 \,, \\
b &= & 1/\sqrt{k} \,, \\
\ga(x) &=& \frac{\Gamma(x)}{\Gamma(1-x)}\,.
\ee
The function
 $P(s)$ is defined for $s$ a non-negative integer as
\be
P(s) = \prod_{n=1}^{s} \ga(n b^2)\,, \qquad \qquad P(0)=1\,,
\label{ps}
\ee
and the coefficients $N_{j_1,j_2,j_3}$ are the $SU(2)_{k-2}$ fusion rules:
\be
N_{j_1,j_2,j_3} =
\left\{
\begin{array}{cl}
1 & \quad \textrm{for} \,\,\,\,\, k-2 \geq j_1+j_2+j_3 \geq \textrm{max}(2j_1,2j_2,2j_3) \\
 & \quad \textrm{and} \,\,\,\,\, j_1 + j_2 + j_3 = 0 \,\,\, \textrm{mod} \,\,2 \\
0 & \quad \textrm{otherwise}
\end{array}
\right. \label{su2fusion} \ee For the \h3\ model, the three-point
functions are~\cite{Teschner:1997ft,Teschner:1999ug}\footnote{We use
the normalization of \cite{Teschner:1999ug}.} (see also
\cite{Giribet:1999ft, Ishibashi:2000fn, Giribet:2000fy, Hosomichi:2000bm, Giribet:2001ft, Hofman:2004ny, Giribet:2004zd})
%\footnote{We use the form of the
%three-point functions of \cite{Teschner:1999ug}, which differs from that in \cite{Teschner:1997ft} by an
%$h$-dependent renormalization of the vertex operators. See appendix \ref{secondap}.}
\be
C_H(h_1,h_2,h_3) = - \frac{b^{1+2b^2}  \up(b)}{2 \pi^2 \ga(1+b^2) } [\nu b^{2b^2}]^{-h+1} \frac{1}{\up(b(h -1))}
\prod_{i=1}^{3} \frac{\up (2 b h_i  -b) }{\up (b(h - 2h_i))} \,,
\label{ch}
\ee
where
\be
h=h_1+h_2+h_3 \,,
\ee
and the function $\up$,
introduced in~\cite{Zamolodchikov:1995aa},
is related to the Barnes double gamma function
and can be defined by
\be
\log \up(x) = \int_0^{\infty}
\frac{dt}{t} \left[\left( \frac{Q}{2} -x \right)^2 e^{-t}
- \frac{\sinh^2((\frac{Q}{2}-x)\frac{t}{2})}{\sinh \frac{bt}{2} \sinh \frac{t}{2b}}
\right]\,.
\ee
The integral converges in the strip $0<\textrm{Re}(x)< Q$.
Outside this range it is defined by the relations
\be
\up(x + b) = b^{1-2bx}\ga(bx) \up(x)
\qquad
\up(x + 1/b) = b^{-1+2x/b}\ga(x/b) \up(x) \,.
\label{shiftu}
\ee
The two-point functions for delta-normalizable states
(with $h=\frac12 + i\rho, \rho \in \mathbb{R}$) can be obtained by
taking one of the operators to be the identity and gives
\be
\langle \Phi_{h= \frac12 + i \rho}(x_1)
 \Phi_{h'= \frac12 + i \rho'}(x_2)
\rangle &=&
\label{h32pf}
\lim_{\epsilon \rightarrow 0} \,\, \langle \Phi_{h=\frac12 + i\rho}(x_1)
 \Phi_{h'= \frac12 + i \rho'}(x_2)
 \Phi_{\epsilon}(x_3)
\rangle  \\
&& \!\!\!\!  \!\!\!\!  \!\!\!\!  \!\!\!\! \!\!\!\! \!\!\!\! \!\!\!\!
= \frac{1}{|z_{12}|^{4 \Delta_h}} \left( \delta^{(2)}(x_1-x_2) \delta(\rho+\rho') + \frac{B(h)}{|x_{12}|^{4h}} \delta(\rho-\rho) \right)
\label{h32pf2}
\ee
where
\be
B(h) = - \frac{\nu^{-2h+1}}{\pi b^2} \ga(1-b^2(2h-1)) \,.
\ee
To obtain  (\ref{h32pf2}) from (\ref{ch}) one should use that
\be
\up(\epsilon) &\sim& \epsilon \up(b) \,,
\ee
which follows from (\ref{shiftu}), and the distributional limits
\be
\lim_{\epsilon \rightarrow 0} \frac{\epsilon}{\rho^2 + \epsilon^2} &=& \pi \delta(\rho) \,,\\
\lim_{\epsilon \rightarrow 0} |x|^{-2+2\epsilon}  &=& \pi \frac{\delta^{(2)}(x)}{\epsilon} \,.
\ee
After taking the limit (\ref{h32pf}), the resulting expression can be analytically
continued to non-normalizable states.
The overall constant in $C_H$ (\ref{ch}) is not determined by the functional
equations of the conformal bootstrap (see appendix \ref{secondap}),
and is fixed by requiring the coefficient of the first term in (\ref{h32pf2}) to be 1.

The parameter $\nu$ is  free in the \h3\ WZW model.
It is not a parameter of the action, which only depends on the level $k$, but rather
of the vertex operators, and
in the conformal bootstrap is left undetermined by the crossing symmetry (see appendix \ref{secondap}).
In the interacting theory, vertex operators
are the sum of an incoming and an outgoing wave, and $\nu$ appears
in the reflection coefficient between these two terms.\footnote{In the free field
realization of the theory, $\nu$ does appear in the action
multiplying the interaction term~\cite{Giveon:1998ns},
because the latter  is used to transform the vertex
operators of the free theory into those of the interacting theory, via insertion of screening charges in the correlators.
The value of $\nu$ is also related to the dilaton in $AdS_3$~\cite{Giveon:2001up}.}
In \cite{Teschner:1999ug} it was proposed to fix $\nu$ by demanding
that the constant
\be
c_{\nu} = \frac{\pi \Ga(1-b^2)}{\nu \Ga(1+b^2)} \,,
\label{nu}
\ee
which appears
in the OPE
\be
\Phi_{1}(x_1) \Phi_{h}(x_2) \sim c_{\nu} \, \delta^{(2)}(x_1-x_2)\Phi_{h}(x_2) \,,
\label{ope}
\ee
be set to $c_{\nu} = 1$
We leave $\nu$ undetermined for the moment, and it will be fixed  below
holographically by comparing the bulk and the boundary correlators.

We will  evaluate the expression (\ref{ch}) for $C_H$
at  values of $h_i$ such that $2h_i$ and $h$ are nonnegative integers.
For these values, eq.(\ref{ch}) can be expressed in terms of  $P(s)$, defined in~(\ref{ps}), by means of the identity
\be
P(s) = \frac{\up(sb +b)}{\up(b)} b^{s((s+1)b^2-1)} \,,
\label{psu}
\ee
which is easily verified by iterating $s$ times the first equation in (\ref{shiftu}).
We get thus
\be
C_H(h_1,h_2,h_3) = -\frac{\nu^{-h+1}  }{2 \pi^2 \ga(1+b^2)}
 \frac{1}{P(h-2)}
\prod_{i=1}^{3}
\frac{P(2h_i-2)}{P(h-2h_i-1)}\,.
\label{chp}
\ee
We are interested in the product
\be
C(h_1,h_2,h_3) \equiv  C_H(h_1,h_2,h_3)
C_S(h_1-1,h_2-1,h_3-1) \,,
\ee
which, from (\ref{zf3pf}) and (\ref{chp}) is equal to
\be
C(h_1,h_2,h_3)=  N_{h_1-1,h_2-1,h_3-1} \frac{\nu^{-h+1} } { 2 \pi^2 b^4 \sqrt{\gamma(b^2)}}
\prod_{i=1}^{3} \frac{1}{\sqrt{\gamma(b^2(2h_i-1))}} \,.
\label{3pp}
\ee
This expression has the remarkable property that the four $P(s)$'s  in $C_S$ and $C_H$ that depend on more than one
of the $h_i$'s have canceled against each other.
In particular,  the poles of $C_H$ that appear at particular linear combinations of several  $h_i$'s,
whose physical meaning was analyzed in \cite{Maldacena:2001km}, have disappeared.
This cancelation of the structure constants follows
from the close relationship between $SU(2)$ and \h3\ structure constants, which we explore further in Appendix B.
Note that the remaining $h_i$-dependent factors can be absorbed by rescaling the $\O_{h_i}$ operators,
as we will do below when normalizing the chiral operators.

Finally,  note that
$\langle \O_{h_1}\O_{h_2}\O_{h_3} \rangle$ is independent of
$z_i,\bz_i$, again due to a cancelation between the dependence on
$z_i,\bz_i$ of the two factors (\ref{3pfphi}) and (\ref{3pfv}).

\subsection{Three-point Correlators of Chiral Multiplets \label{Three}}

Now that we have the correlation function of three $\O_h(x,y)$'s, in order to compute
the three-point function of the chiral $SU(2)$ multiplet  we only need to add factors involving
the fermions, spin fields and current algebra descendants.
To illustrate the steps involved,
let us compute in detail the following correlator
\be
\langle  c\bar{c}\O_{h_1}^{(0,0)}  c\bar{c}\O_{h_2}^{(0,0)} c\bar{c}\tilde{\O}_{h_3}^{(0,0)}  \rangle \,,
\ee
where the pictures have been chosen so that the total picture number adds up to $-2$.
This correlator will  be proportional to $C(h_1,h_2,h_3)$,
but we are interested in the precise prefactor.
The last term in (\ref{zppo}) of the zero picture vertex $\tilde{\O}_{h_3}^{(0,0)}$ can
be discarded in this particular correlator since $\langle \chi_a \rangle = \langle \bar{\chi}_a \rangle =0$.
%\be
%&&
%\frac{2}{k} \big \langle  \bar{\psi}(\bx_1)\bar{\chi}(\by_2) \bar{\psi}(\bx_3) \bar{\chi}_a P^a_{\by_3} \big \rangle
%\times \\
%&&  \qquad  \times \,\,  \Big[
%\big  (1-h_3) \langle
%     \psi(x_1) \psi(x_2)   \hat{\jmath}(x_3)  \big \rangle
%\big \langle
%\O_{h_1}
%\O_{h_2}
%\O_{h_3}
%\big \rangle +
% \nn
%%&& \qquad  +
% \big  \langle \psi(x_1)\psi(x_2)  \big \rangle
%%\big \langle  \frac{2}{k} \bar{\psi}(\bx_1)\bar{\chi}(\by_2) \bar{\psi}(\bx_3) \bar{\chi}_a P^a_{\by_3} \big \rangle
%\big \langle
%\O_{h_1}
%\O_{h_2}
%j(x_3) \O_{h_3}
%\big \rangle \Big]
%\ee
For the computation we need to use
\be
\langle {\psi} (x_1; z_1)
{\psi}(x_2; z_2) \rangle  = k\frac{(x_{12})^2}{z_{12}}
\,, \label{2pfpsi}
\ee
and
\be
\langle \psi(x_1;z_1) \psi(x_2;z_2) \hat{\jmath}(x_3;z_3)\rangle
= - 2k\frac{x_{12}x_{23} x_{31}} {z_{13}z_{23}}\,,
\ee
which follows from
\be \langle \psi^A(z_1) \psi^B(z_2) \hat{\jmath}^C(z_3) \rangle
= \frac{i\frac{k}{2}\ve^{ABC}}{z_{13}z_{23}} \,.
\ee
We also need the value of
$\langle \O_{h_1}\O_{h_2}j(x_3)\O_{h_3} \rangle$, where normal order implies
\be
j(x_3) \O_{h_3}(x_3)= \left( -j_{-1}^{+} +2x_3j_{-1}^3 - x_3^2j_{-1}^{-} \right) \O_{h_3}(x_3) \,.
\ee
The correlation
functions of these current algebra descendants can  be
expressed in terms of correlators of the primaries by combining the
Ward identity
\be
\langle j^A(w) \Phi_{h_1}(x_1; z_1) \ldots
\Phi_{h_3}(x_3; z_3) \rangle = - \sum_{i=1}^3 \frac{D^A_{x_i} }
{w-z_i} \langle \Phi_{h_1}(x_1;z_1) \ldots \Phi_{h_3}(x_3; z_3)
\rangle \label{ward}
\ee
with the OPE
\be j^A(w)\Phi_{h_3}(x_3 ;z_3) \sim \!-\!\frac{D_{x_3}^a}{w-z_3} \Phi_{h_3}(x_3;z_3) \!+\! j_{-1}^A
\Phi_{h_3}(x_3;z_3) \!+\! j_{-2} ^A \Phi_{h_3}(x_3 ; z_3) (w\!-\!z_3)\!+\!\cdots  \label{opecurrent}
\ee
Expanding the $i=1,2$
denominators in (\ref{ward}) as $(w-z_i)^{-1} =
(z_3-z_i)^{-1}\sum_{n=0}^{\infty} \left(
\frac{w-z_3}{z_i-z_3}\right)^{n}$, eqs.(\ref{ward})
and~(\ref{opecurrent}) give
\be
\langle \O_{h_1}(x_1)
 \O_{h_2}(x_2)
 j^A_{-1}\O_{h_3}(x_3)
 \rangle
 = \left( \frac{D^A_{x_1}}{z_{13}} + \frac{D^A_{x_2}}{z_{23}} \right)
\langle \O_{h_1}(x_1)
 \O_{h_2}(x_2) \O_{h_3}(x_3) \
\rangle \,,
\ee
and using now the dependence of $\langle \O_{h_1}\O_{h_2}\O_{h_3} \rangle$  on $x_i$ given
by (\ref{3pfphi}), we obtain
\be
\label{obtain}
\langle \O_{h_1}
 \O_{h_2} j(x_3)\O_{h_3}  \rangle =
(-h_1-h_2+h_3)
\frac{z_{12}} {z_{31}z_{23}}
\frac{x_{23}x_{31} }{x_{12}}
\langle \O_{h_1}
 \O_{h_2} \O_{h_3} \rangle \,.
\ee
Note that although $\langle \O_{h_1}\O_{h_2}\O_{h_3} \rangle$ is
independent of $z_i,\bz_i$, the above expression does depend on
$z_i$.
Using
\be
\langle c(z_1)c(z_2)c(z_3) \rangle = z_{12}z_{23}z_{31}\,,
\ee
collecting all the terms, and including the anti-holomorphic factors, we get finally
\be
\langle
c\bar{c}\O_{h_1}^{(0,0)}  c\bar{c}\O_{h_2}^{(0,0)}
c\bar{c}\tilde{\O}_{h_3}^{(0,0)}   \rangle  &=&
 g_s v_4 k^2  (h_1+h_2+h_3-2)^2 \,\,
 C(h_1,h_2,h_3)
\label{3pfm}
 \\
 \nn
&& \times \,\,
\left|\frac{y_{12}} {x_{12}}\right|^{2H_1+2H_2-2H_3}
\left|\frac{y_{23}} {x_{23}}\right|^{2H_2+2H_3-2H_1}
\left|\frac{y_{31}} {x_{31}}\right|^{2H_3+2H_1-2H_2}
\ee
\vskip 0.5cm
\ni
where we have included the correct power of $g_s= g_s^{-2+3}$ and the volume $v_4$ of the $T^4$.
The other correlators are computed similarly, and in all the cases
holomorphic and anti-holorphic
contributions factorize. Labeling the operators with  $\a,\bar{\a}=0,a,2$, all the correlators
have the form (we omit the dependence on $x_i, y_i$, which is the same as above)
\be
\langle
c\bar{c}\O_{h_1}^{(\a_1,\bar{\a}_1)}  c\bar{c}\O_{h_2}^{(\a_2,\bar{\a}_2)} c\bar{c}\tilde{\O}_{h_3}^{(\a_3,\bar{\a}_3)} \rangle
&=& g_s v_4 \,    C(h_1,h_2,h_3) \, g(h_i;\a_1,\a_2,\a_3) g(h_i;\bar{\a}_1,\bar{\a}_2,\bar{\a}_3)\,
\ee
with
\be
g(h_i;0,0,0) &=& k(-h_1 -h_2 -h_3 +2) \,,
\label{3pffirst}
\\
g(h_i;0,0,2) &=& k(-h_1-h_2 + h_3 +1) \,,
\\
g(h_i;0,2,2) &=& k(-h_1 +h_2+h_3) \,,
\\
g(h_i;2,2,2) &=& k(h_1 + h_2 +h_3 -1) \,,
\\
g(h_i;0,a,b) &=& \sqrt{k} \xi^{ab} \,,
\\
g(h_i;2,a,b) &=& \sqrt{k} \xi^{ab} \,,
\label{3pflast}
\ee
%\be
%\langle
%c\bar{c}\O_{h_1}^{(0,0)}  c\bar{c}\O_{h_2}^{(0,0)}
%c\bar{c}\tilde{\O}_{h_3}^{(2,2)}   \rangle
%&=& g_s v_4 \,
%  k^2  (h_1+h_2- h_3-1)^2 \,\,  C(h_1,h_2,h_3)
%\\
%\langle
%c\bar{c}\O_{h_1}^{(0,0)}  c\bar{c}\O_{h_2}^{(2,2)}
%c\bar{c}\tilde{\O}_{h_3}^{(2,2)}   \rangle
%&=& g_s v_4 \,
%  k^2  (h_2+h_3-h_1)^2  \,\, C(h_1,h_2,h_3)
%\\
%\langle
%c\bar{c}\O_{h_1}^{(2,2)}  c\bar{c}\O_{h_2}^{(2,2)}
%c\bar{c}\tilde{\O}_{h_3}^{(2,2)}   \rangle
%&=& g_s v_4 \,
%  k^2  (h_1+h_2+h_3-1)^2 \, \, C(h_1,h_2,h_3)
%\\
%\langle
%c\bar{c}\O_{h_1}^{(a,\bar{a})}  c\bar{c}\O_{h_2}^{(b,\bar{b})}
%c\bar{c}\tilde{\O}_{h_3}^{(0,0)}   \rangle
%&=& g_s v_4 \, \xi^{a\,b}  \xi^{\bar{a}\,\bar{b}} \,
%  k \,\,  C(h_1,h_2,h_3)
%\\
%\langle
%c\bar{c}\O_{h_1}^{(a,\bar{a})}  c\bar{c}\O_{h_2}^{(b,\bar{b})}
%c\bar{c}\tilde{\O}_{h_3}^{(2,2)}   \rangle
%&=& g_s v_4 \, \xi^{a\,b}  \xi^{\bar{a}\,\bar{b}} \,
%  k \, \, C(h_1,h_2,h_3)
%\label{3pflast}
%\ee

\vskip 1cm
\ni
 and
the results are independent of the picture chosen for the  operators,
as long as the total picture number is~$-2$.
%The structure constants in the chiral ring are obtained from the three point function
%of two chiral operators and one antichiral. (To be followed...)

%Imposing $M_i=J_i$ is equivalent to

%\be
%\langle S^-(x_1,y_1) S^-(x_2, y_2) \psi(x_3) \rangle  = \\
%\langle S^-(x_1,y_1) S^-(x_2, y_2) \chi(y_3) \rangle  = \\
%\ee

\subsection{Two-point Functions}
In order to compare the three-point functions of the bulk to those of the symmetric product orbifold,
operators of both sides should be equally normalized.
In the symmetric product the normalization is given by (\ref{boundary2pf})-(\ref{boundary2pf2}).
To compute a two-point function in the string theory side, when fixing two
vertex operators in the sphere there is a zero coming  from dividing by the volume of the conformal
group. This zero is canceled against the divergence of the delta~$\delta(h_1-h_2)$ in the two-point function in the \h3\
 WZW model (\ref{h32pf2}), which can be interpreted as
the volume of the Killing group in the target space which leaves invariant
the positions $x_1,x_2$ of the two operators \cite{Kutasov:1999xu}.
As we review now,  the finite result of this cancellation is $h$-dependent.

The string two-point function can be obtained, following \cite{Maldacena:2001km} (see  also~\cite{Aharony:2003vk}),
by exploiting the Ward identity for affine currents in the boundary CFT.
Given an holomorphic affine current~$K^a(z)$ in the inner CFT of an $AdS_3$ compactification,
it was shown in \cite{Kutasov:1999xu} that the vertex operator
\be
K^a(x;z) =  - \frac{1}{k c_{\nu}} \, K^a(z) \bar{j}(\bx) \Phi_{1}(x,\bx)
\label{curr}
\ee
is the corresponding holomorphic affine current in the dual CFT, with $c_{\nu}$
defined in (\ref{nu})-(\ref{ope}).
Note that it has the correct conformal dimensions $(\Delta, \bar{\Delta})= (1,1)$ in the bulk,
and $(H, \bar{H})=(1,0)$ in the boundary.

The Ward identity for the above current in  the boundary CFT is
\be
\langle c\bar{c}K^a(x) \Phi_h(x_1)P_1 \,\,  \Phi_h(x_2) P_2 \rangle
=
\left(\frac{q_1}{x-x_1} + \frac{q_2}{x-x_2} \right) \langle \Phi_h(x_1)P_1 \,\,  \Phi_h(x_2) P_2 \rangle
\label{wardbound}
\ee
where the  $P_{1,2}$ stand for  the ghosts, fermions and operators of the internal theory,
and $q_{1} = -q_{2}$ are the charges of $P_{1,2}$ under $K^a(z)$.
Note that the expression  for $K^a(x;z)$ in (\ref{curr}) is
in the zero picture, so that on both sides of (\ref{wardbound}) the operators $\Phi_h(x_i)P_i \,\, (i=1,2)$ are in the same picture.
The lhs of (\ref{wardbound}) can be computed as we did in the previous subsection (see eq.(\ref{obtain})).
Comparing the resulting expression with the rhs yields the string theory two-point function
\be
\langle \Phi_h(x_1)P_1 \,\,  \Phi_h(x_2) P_2 \rangle =
- \frac{1}{k c_{\nu}}
\frac{(2h-1)C_H(1,h,h)\, p_{12}}{|x_{12}|^{4h} }  \,,
\label{string2pf}
\ee
where $p_{12}$ is
\be
p_{12} =  \langle c(\infty) \bar{c}(\infty) P_1(1) P_2(0)\rangle \,.
%p_{12} = \frac{1}{|z_{12}|^2 |z_{23}|^2 |z_{31}|^2 } \frac{\langle c(z_3) \bar{c}(\bz_3) P_1 P_2\rangle}{|z_{12}|^{4\Delta_h-4}}
\ee
There are no powers of $g_s$ for the two-point functions.
Note that
\be
C(1,h,h) = \frac{b^2\ga(-b^2)}{2 \pi \nu} B(h)\,,
\ee
so the difference between the string theory two-point function (\ref{string2pf}) and  that of the \h3\ WZW model (\ref{h32pf2})
is the factor $(2h-1)$, plus $h$-independent factors.

Choosing the chiral operators
so that the total picture is $-2$, we get, for the scalar sector,
\be
\langle c\bar{c}\O^{(0,0)}_{h} c \bar{c}\O^{(0,0)}_{h} \rangle =
\langle c\bar{c}\O^{(2,2)}_{h} c\bar{c}\O^{(2,2)}_{h} \rangle &=&
 -c^{-1}_{\nu} k (2h-1) C(1,h,h) v_4 \, {\left| \frac{y_{12}}{x_{12}}\right|}^{4H} \\
\langle c\bar{c}\tilde{\O}^{(a,\bar{a})}_{h} c\bar{c}\O^{(b,\bar{b})}_{h} \rangle &=& - c_{\nu}^{-1}  (2h-1)^{-1}
C(1,h,h) v_4  \, \xi^{a\,b}  \xi^{\bar{a}\,\bar{b}} \,
{\left| \frac{y_{12}}{x_{12}}\right|}^{4H}
\ee
where we used $C_H(1,h,h)=C(1,h,h)$.
In the  last line
we have taken into account the prefactor carried by the operators in the $-3/2$ picture (see (\ref{th})).

%and similarly
%\be
%\langle c\bar{c}\O^{(2,2)}_{h_i} c\bar{c}\O^{(2,2)}_{h_j} \rangle
%&=& \nu^{-2h-1} b^{-2 + 2b^2} \,
%\frac{\gamma(b^2)}{\up(b)} \frac{ \ga(b^2(2h-1))(2h-1)  |y_{12}|^{4J}}{|x_{12}|^{4H}}
%\\
%\langle c\bar{c}\tilde{\O}^{(a,a)}_{h_i} c\bar{c}\O^{(b,b)}_{h_j} \rangle &=&
%\nu^{-2h-1}
%b^{2+2b^2} \,
%\frac{\gamma(b^2)}{\up(b)} \frac{ \ga(b^2(2h-1))(2h-1)^{-1}  |y_{12}|^{4J}}{|x_{12}|^{4H}}
%\ee
The (c,c) elements of the $N=2$ chiral ring  are those operators with $M=\bar{M}=J$, where $M,\bar{M}$ are the eigenvalues
of $K^3_0,\bar{K}_0^3$. Similarly, the (a,a) operators correspond to  $M=\bar{M}=-J$.
Therefore we define the normalized (c,c)  operators as (see the expansion~(\ref{yexpansion}))
\be
\mathbb{O}_h^{(0,0)} &=& c\bar{c}\O_h^{(0,0)}(y=\bar{y}=0) \,\,
  \left[ -c^{-1}_{\nu} k  (2h-1) C(1,h,h)v_4 \right]^{-1/2} \,,
\label{2pfo}
\\
\mathbb{O}_h^{(2,2)} &=& c\bar{c}\O_h^{(2,2)}(y=\bar{y}=0)
\, \,  \left[ -c^{-1}_{\nu} k (2h-1) C(1,h,h)v_4 \right]^{-1/2} \,,
\\
\mathbb{O}_h^{(a,a)} &=& c \bar{c} \O_h^{(a,a)}(y=\bar{y}=0)
 \left[ -c^{-1}_{\nu}  (2h-1)^{-1} C(1,h,h) v_4\right]^{-1/2} \,,
\ee
and the (a,a) operators as
\be
\mathbb{O}_h^{(0,0)}{}^{\dagger} &=& \lim_{y,\bar{y} \rightarrow \infty} |y|^{-4H} c\bar{c}\O_h^{(0,0)}(y,\bar{y})  \,\,
 \left[ -c^{-1}_{\nu} k  (2h-1) C(1,h,h) v_4 \right]^{-1/2} \,,
\label{2pfod}
\ee
and similarly for $\mathbb{O}_h^{(2,2)}{}^{\dagger}$ and $\mathbb{O}_h^{(a,a)}{}^{\dagger}$.
Note that we have included the ghosts $c\bar{c}$ in the definition.
These operators are thus normalized as
\be
\langle \mathbb{O}_h^{(0,0)}{}^{\dagger} \mathbb{O}_h^{(0,0)} \rangle
&=&
\langle \mathbb{O}_h^{(2,2)}{}^{\dagger} \mathbb{O}_h^{(2,2)} \rangle
= \frac{1}{|x_{12}|^{4H}} \\
\langle \tilde{\mathbb{O}}_h^{(a,\bar{a})}{}^{\dagger} \mathbb{O}_h^{(b,\bar{b})} \rangle &=&
\frac{\delta^{a \, b} \delta^{\bar{a} \, \bar{b}}}{|x_{12}|^{4H}}
\ee
One can define similarly normalized operators for the non-scalar sector.

\subsection{Fusion Rules}
Before computing the structure constants, let us see how the boundary
fusion rules (\ref{fusionrules}) are obtained in the bulk.
%We prove this using the $SU(2)_{k-2}$ fusion rules.
The chiral (anti-chiral) operator in each $SU(2)$ multiplet corresponds to $M=J$ ($M=-J$).
Therefore, conservation of the $U(1)$ R-charge, measured with $K^3_0$, implies that
the fusion
\be
\O_{h_1}^* \times \O_{h_2}^* = [\O_{h_3}^*]
\ee
is possible only if
\be
J_3 =  J_1 + J_2 \,.
\label{u1cons}
\ee
On the other hand,
\be
J_i = j_i + \hat{\jmath}_i
\ee
where $\hat{\jmath}_i = 0,1/2,1$ for $\O^{0}, \O^{a}, \O^{2} $, respectively.
Now, from the fusion rules (\ref{su2fusion}) it follows that
\be
j_3 \leq j_1 + j_2
\label{litjin}
\ee
and therefore (\ref{u1cons})
implies\footnote{The inequality (\ref{hatjin}) does not mean
that there is a violation of the rules of $SU(2)$ tensor product for the $\hat{k}^a$ algebra.
Since the operators apperar in different pictures, the label $\hat{\jmath}_i$
here only denotes the family to which an operator belongs, but not necessarily its spin under $\hat{k}^a$.}
%\footnote{Note that the inequalities  in (\ref{litjin}) and
%(\ref{hatjin}) are not in contradiction with the conservation of  $j^3_0$ and $\hat{\jmath}^3_0$.}
\be
\hat{\jmath}_3 \geq  \hat{\jmath}_1 + \hat{\jmath}_2 \,.
\label{hatjin}
\ee
The equality corresponds to the four cases
\be
(0) \times (0) &=& (0)  \nn \\
(0) \times (2) &=& (2) \label{idenfusion} \\
(0) \times (a) &=& (a) \nn \\
(a) \times (a) &=& (2) \nn
\ee
and
the inequality to the case
\be
(0) \times (0) &=& (2).
\label{ineqfusion}
\ee
Note that operators of the Ramond sector appear in pairs.
For the antiholomorphic quantum numbers we get similarly
\be
\bar{\hat{\jmath}}_3 &\geq & \bar{\hat{\jmath}}_1 + \bar{\hat{\jmath}}_2 \,.
\label{hatjina}
\ee
For  chiral operators we should combine both rules (\ref{hatjin}) and (\ref{hatjina}),
taking into account that the $j_i$ quantum numbers are
the same for holomorphic and anti-holomorphic part. This implies that
the four cases (\ref{idenfusion}) can be freely combined between
holomorphic and anti-holomorphic, and the case (\ref{ineqfusion}) should
be the same in the holomorphic and anti-holomorphic sectors.
These are precisely the same fusion rules as in the boundary theory,
associated to the same $1 + 4\times 4= 17$ processes.

\subsection{Structure Constants}

We have already  all the elements to compute the structure constants of the chiral ring.
Consider first $\langle \mathbb{O}_{h_3}^{(0,0)}{}^{\dagger}  \mathbb{O}_{h_2}^{(0,0)}
\tilde{\mathbb{O}}_{h_1}^{(0,0)} \rangle$.
From $H_3=H_1+H_2$ we get $h_3 = h_1+h_2-1$.
Plugging this into (\ref{3pfm}) and using~(\ref{3pp}),~(\ref{2pfo})  and~(\ref{2pfod}) gives
\be
\langle \mathbb{O}_{h_3}^{(0,0) \,\dagger }  \mathbb{O}_{h_2}^{(0,0)} \tilde{\mathbb{O}}_{h_1}^{(0,0)} \rangle
&=&
\frac{g_s}{\sqrt{v_4}} \left(\frac{2 \pi }{\nu \ga(1+b^2)}\right)^{1/2} \, \left[\frac{(2h_3-1)^3}{(2h_1-1)(2h_2-1)} \right]^{1/2}
%\\
%&=& \left(\frac{1}{N}\right)^{1/2} \, \left[\frac{(2h_1+2h_2-3)^3}{(2h_1-1)(2h_2-1)} \right]^{1/2}
\label{ooo}
\ee
where we have fixed $x_1=\bar{x}_1=0, x_2=\bar{x}_2=1,  x_3=\bar{x}_3=\infty$.
In order to get the  prefactor $N^{-1/2}$ to agree with (\ref{boundary3pf1}), we use that
\begin{equation}
\label{N}
    \frac{1}{\sqrt{N}} = \frac{1}{\sqrt{Q_1 Q_5}} = \sqrt{\frac{Q_5}{Q_1}} \frac{1}{Q_5} = g_6 \frac{1}{Q_5}= \frac{g_s}{\sqrt{v_4}} \frac{1}{Q_5} \,,
\end{equation}
and this fixes the value of $\nu$,
which was the only free parameter in the \h3\ WZW model, as
\be
\nu = \frac{2\pi}{b^4 \ga(1+b^2)} \,.
\label{fixnu}
\ee
We are considering  fixed $Q_5$ but large $Q_1$
so that the string coupling is small and $N$ is large.
With this value for $\nu$, the other four scalar correlators are computed similarly and are
\be
\langle \mathbb{O}_{h_3}^{(2,2) \,\dagger}  \mathbb{O}_{h_2}^{(0,0)} \tilde{\mathbb{O}}_{h_1}^{(0,0)} \rangle
&=&
\left(\frac{1}{N}\right)^{1/2}
\left[\frac{1}{(2h_1-1)(2h_2-1) (2h_3-1)} \right]^{1/2}
\\
\langle \mathbb{O}_{h_3}^{(2,2)\,\dagger}  \mathbb{O}_{h_2}^{(0,0)} \tilde{\mathbb{O}}_{h_1}^{(2,2)} \rangle
&=&
\left(\frac{1}{N}\right)^{1/2}
\left[\frac{(2h_1-1)^3}{(2h_2-1)(2h_3-1)} \right]^{1/2}
\\
\langle \mathbb{O}_{h_3}^{(a,\bar{a}) \, \dagger}  \mathbb{O}_{h_2}^{(0,0)} \mathbb{O}_{h_1}^{(b,\bar{b})} \rangle
&=&
\left(\frac{1}{N}\right)^{1/2}
\left[\frac{(2h_1-1)(2h_3-1)}{(2h_2-1)} \right]^{1/2} \delta^{a \, b} \delta^{\bar{a} \, \bar{b}}
\\
\langle \mathbb{O}_{h_3}^{(2,2) \, \dagger}  \mathbb{O}_{h_2}^{(a,\bar{a})} \mathbb{O}_{h_1}^{(b,\bar{b})} \rangle
&=&
\left(\frac{1}{N}\right)^{1/2}
\left[\frac{(2h_1-1)(2h_2-1)}{(2h_3-1)} \right]^{1/2}  \xi^{a\,b}  \xi^{\bar{a}\,\bar{b}} \,,
\ee
where in all the cases $h_3$ is a obtained from  $h_{1,2}$  using $H_3=H_1+H_2$.
All correlators coincide precisely with the boundary results (\ref{boundary3pf1})-(\ref{boundary3pf5}).
It is immediate to extend the agreement to the  12 correlators of the non-scalar sector,
using the factorization of the structure constants
into holomorphic and anti-holomorphic contributions,
which holds both in the bulk and in the boundary.
Much like in the boundary, it is easy to see that
the (c,a) ring in the bulk has the same fusion rules and
structure constants as the (c,c) ring.

Contrary to the boundary correlators (\ref{boundary3pf1})-(\ref{boundary3pf4}), the above bulk correlators
are defined for $\mathbb{O}_{h_2=1}^{(0,0)}$, but, as we mentioned above, the three-point functions
do not reduce to the two-point functions as would be for an identity operator.

Given the different normalizations of the $\mathbb{O}_h$ operators and the
different powers of $k$ in (\ref{3pffirst})-(\ref{3pflast}), it is  remarkable  that the definition (\ref{fixnu})
gives the correct prefactor in all the  cases.

Note that the volume $v_4$ of the inner $T^4$ disappears from the correlators and from $\nu$,
and we could have used  $g_6$ from the beginning.
This is consistent with the supergravity result that in the frame with NS-NS flux, the value of $v_4$ is an arbitrary number
not related to the value of $T^4$ in the boundary or to other parameters of the theory.

\section{Discussion  \label{Conclude}}

The remarkable agreement between bulk and boundary quantities
computed at different points in moduli space begs for an explanation.
A similar agreement between the three-point correlators of $N=4$
super Yang-Mills theory and the dual  $AdS_5 \times S^5$
case \cite{Freedman:1998tz,Lee:1998bx} was explained by showing
that two- and three-point functions do not receive $g^2_{YM}$ corrections
\cite{D'Hoker:1998tz,Skiba:1999im, Penati:2000zv, Gonzalez-Rey:1999ih, Intriligator:1998ig, Intriligator:1999ff,Howe:1998zi}.
We conjecture that a similar non-renormalization exists in our case,
which should be further investigated.

Another possibility, if we accept the model of the iterated structure (\ref{iterated}), is that
the $Z_2$ twists which deform the orbifold \cite{David:1999ec} to
the point where the bulk string theory has NS-NS flux only, are such that they only mix
different $Q_1$ copies but do not mix the $Q_5$ copies, and therefore
the deformation is not seen when considering unflowed $SL(2,R)$ representations
in the bulk. To settle this question definitively more work is needed. In particular it would
be useful to compute the correlators of spectrally flowed operators in the bulk as
well as the correlators in the iterated symmetric product in the boundary.
These computations would hopefully provide enough additional information to arrive
at the correct interpretation. We plan to return to these questions
in \cite{Dabholkar:2006nn}.

In other $AdS_{n+1}/CFT_n$ backgrounds with $RR$ fields,
bulk computations are mostly limited to supergravity and
it is not clear how to even begin the computation of  loop amplitudes.
An important advantage of the $AdS_3/CFT_2$  background with NS-NS flux considered here
is that one has an exact worldsheet
description available for the bulk string theory.
It is natural to ask
if the striking agreement found at tree level extends to higher loops.
Fortunately, exact answers for finite $N$ are available in the boundary.
In the bulk,  quantum corrections to  three-point correlators can in principle be computed
systematically by evaluating higher genus string amplitudes.
It would be very interesting to see if the technical tools can be developed
sufficiently to carry out such a comparison.
We hope our results and further investigations will lead to a better
understanding of the chiral sector  of the theory for finite $N$ and
also to more stringent tests of the gauge-string duality.

%In our case,  the iterated structure (\ref{iterated}) suggests
%that the deformation in the boundary off the orbifold point, to the point corresponding
%to the NS-NS background in the bulk,  leaves the $\textrm{Sym}^{Q_5}$ orbifold
%intact and only deforms the $\textrm{Sym}^{Q_1}$ structure. Such a deformation would not be seen in
%our computations, which are performed in the untwisted sector of $\textrm{Sym}^{Q_1}$,
% and this would explain the precise agreement. The possible deformations off the
%orbifold point are $Z_2$ twists \cite{David:1999ec}, which therefore should  act leaving the
%untwisted sector of  $\textrm{Sym}^{Q_1}$ invariant.\footnote{These $Z_2$ twists
%are similar to the DVV operator in matrix string theory \cite{Dijkgraaf:1997vv}.} It would be interesting
%to explore the interactions on the deformed orbifold and compare them to correlators
%of the spectrally flowed sectors.\footnote{An interesting problem is that one
%should be able to reproduce  from the boundary the
%non-conservation of the spectral flow number by up to one unit that occurs in the three-point
%function of the $H_3^+$ WZW model~\cite{FZZ,Maldacena:2001km}.}

%In the bulk, it would be interesting to see if, as we expect, the chiral correlators
%are left invariant when turning on a RR field using the formalism developed in  \cite{Berkovits:1999im}
%(see also the  recent work \cite{Gotz:2006qp}).

The cubic  couplings of chiral primaries in this background
have been studied in the supergravity limit
in \cite{Mihailescu:1999cj,Arutyunov:2000by,Donos:2005vs,Kanitscheider:2006zf},
but no agreement was found with the boundary results.
We believe our results might help to better understand those computations.
%In particular, we do not expect that  supergravity can give the fusion rules of the chiral ring.
%The reason is that in the
%worldsheet the latter  come  from
%the fusion rules of the $SU(2)$ WZW model, which in turn follow from the
%existence of degenerate operators in the $SU(2)_k$ affine algebra \cite{Zamolodchikov:1986bd, Gepner:1986wi},
%but only the global part of that algebra survives in the supergravity limit.

More generally, it has been pointed out in the past  that
some aspects of holography in  this
background follow a paradigm close to the
matrix models duals of non-critical strings~\cite{Martinec:1998st}.
The cancelations between the three-point functions,
similar to the minimal strings case, strengthen this idea.
Note that in our case the holographic correspondence does not involve legpole factors.
Moreover, a ground ring exists also in our background
 \cite{Rastelli:2005ph},\footnote{The existence of a ground ring
in $AdS_3$ backgrounds is not in contradiction with the vanishing theorem shown to hold in \cite{Pakman:2003kh, Asano:2003qb},
which states  that the cohomology of strings in $AdS_3$ backgrounds is concentrated at ghost number one (except for a few states at ghost number zero).
The reason is that the vanishing theorem assumes that the $SL(2,R)$ spin satisfies the bound (\ref{hboundf}),
and the ground ring elements violate this bound.}
and much like for non-critical strings, it  might lead to an integrable structure shared
by the two holographic descriptions.

%Exploiting the properties of the ground ring one could
%study higher point functions by generalizing the techniques of \cite{Seiberg:2003nm,Belavin:2005af,Kostov:2005av}.

\vskip 1cm
\noindent
{\bf Note added}: upon completion of this work, we learnt of the preprint \cite{Gaberdiel:2007vu},
where one of the five families of correlators discussed here has been computed independently.

%It would be interesting to compute four-point functions.
%Even though a direct computation seems technically challenging,

\vskip 1.2cm

\section*{Acknowledgements}

We thank Rajesh Gopakumar, Amit Giveon, Nori Iizuka, Antal Jevicki, Juan Maldacena, Samir Mathur, Shiraz Minwalla,
Emil Martinec, Sameer Murthy, Andrei Parnachev, Sanjaye Ramgoolam, Leonardo Rastelli, David Sahakyan, Nathan Seiberg, Amit Sever,
Kostas Skenderis, Marika Taylor, Joerg Teschner, Jan Troost and  Cumrun Vafa    for useful conversations and correspondence.
This work started while AP was a Visiting Fellow at the Tata Institute for Fundamental Research.
AD thanks the Institute for Advanced Study, ASICTP and
the Aspen Center for Physics for hospitality. AP thanks Harish-Chandra Research Institute and Brown
University for hospitality. Both authors thank Harvard University for hospitality and the organizers
of the Simons Workshop in Mathematics and Physics 2006. The work of AP is supported by the Simons Foundation.

\newpage

\appendix

\section{Derivation of the  Chiral Spectrum in the Bulk \label{firstap}}
In this appendix we will compute
the spectrum of chiral
operators of the bulk theory \cite{Kutasov:1998zh}
in the $x,\bx$ basis.
Chiral operators
belong to $SU(2)$ multiplets satisfying
\be H=J\,. \label{HJ}
\ee
From~(\ref{jsplit}) and~(\ref{ksplit}),
we see that the representations of $J^A,K^a$ arise as tensor
products of those of the bosonic currents $j^A,k^a$ and the
fermionic currents $\hat{\jmath}^A,\hat{k}^a$. The latter have
representations of spins~$\hat{h},\hat{\jmath}=0,1$ in the NS
sector, and $\hat{h}=\hat{\jmath}=1/2$ in the R sector.

Since $2J \in \mathbb{Z}^+$, it follows from (\ref{HJ}) that the \slr\
factor of a chiral operator will belong to a discrete
representation. States in the Hilbert space of the bosonic
\slr\ and $SU(2)$ WZW models should then
satisfy the bounds (\ref{hboundf}) and (\ref{jboundf}),
\be
\frac12 < h <
\frac{k+1}{2}\,,
\label{hbound}
\\
 0 \leq j \leq  \frac{k-2}{2} \,.
 \label{jbound}
\ee
A physical operator should also
be BRST invariant and survive the GSO projection. We will consider
first the holomorphic spectrum, and afterwards we will discuss its
tensoring with the  anti-holomorphic sector.

\subsection{Neveu-Schwarz Sector \label{NSsector}}

In this sector, commutation with the BRST charge implies two conditions.
Firstly,  the vertex operator
must be a Virasoro primary satisfying the mass shell condition.
In the  $-1$ picture this is
\be
\Delta = -\frac{h(h-1)}{k} + \frac{j(j+1)}{k}
 +\frac{\hat{h}(\hat{h}+1)}{4}
 + \frac{\hat{\jmath}(\hat{\jmath}+1)}{4} + \Delta_T + N = \frac12 \,,
\ee
where $\Delta_T \geq 0$ corresponds to a primary of $T^4$ or $K3$ that
may appear in the vertex operator, and $N$ is the level of
possible excited states. Secondly, there should be no
double poles in the OPE between the vertex operator and the supercurrent $T_F$.
Let us consider the four different $\hat{h}, \hat{j}=0,1$ cases:
%\be
%P_N(\psi, \chi, \dots) \Phi_h V_j  T
%\ee
%where $T$ is a primary field of $T^4/K3$.

\begin{enumerate}
\item $\hat{h}= \hat{\jmath} =0$
\\
In this case $H=h=j=J$, and the mass shell condition is
\be \Delta=
\frac{2h}{k} + \Delta_T + N = \frac12\,.
\ee
Since there are no fermions from \sl\ or $SU(2)$, in order to survive the
GSO projection a fermion from the $\CM$ factor should be excited,
with $\Delta_T=1/2$. This implies $h=0$, which is forbidden by the
bound (\ref{hbound}) on $h$. Thus there are no physical chiral
operators in this sector.

\item $\hat{h}=1, \, \hat{\jmath}= 0  $ \\
In this case we have $J=j$, and the tensor product of $\hat{h}=1$
with $h$ gives the representations $H=h-1,h,h+1$. Let us consider each one of the cases.

\begin{enumerate}
\item $H=h+1 = j=J$ \\
In this case the mass shell condition is
\be
\Delta = \frac{4h+2}{k} + \frac12 + \Delta_T +N =  \frac12
\ee
and would require $h\leq -1/2$, which violates the bound (\ref{hbound}) on $h$,
so there are no physical chiral operators from this sector.

\item $H=h = j=J$
\\
The mass shell condition is
\be
\Delta = \frac{2h}{k} + \frac12 + \Delta_T +N =  \frac12
\ee
and would require $h\leq 0$, which violates the bound (\ref{hbound}) on $h$,
so there are no physical chiral operators from this sector either.

\item $H=h -1 = j=J$
\\
The mass shell condition is \be \Delta = 0 +  \frac12 + \Delta_T +N
=  \frac12 \ee so it is satisfied by $\Delta_T=N= 0$. In section
\ref{Tensor} we work out the details of the $h-1$ representation
coming from the tensor product of the \slr\ bosonic representation
of spin $h$ and the fermionic representation of spin $1$. The result
in the $x$-basis is just \be \Phi_{h} (x)  \psi(x) \,,
\label{phipsi} \ee where $\psi(x)$ is the fermionic \slr\ primary
with $h=-1$ defined in (\ref{psidef}). The product of $\psi(x)$ with
$\Phi_{h}(x)$ has no singularities, since $\Phi_h(x)$ is a primary
of the purely bosonic currents $j^A$.
Let us  define the operator
\be
{\cal O}_h(x,y) \equiv  \Phi_{h} (x) V_{h-1}(y)\,,
\label{odef}
\ee
and note
that \be \Delta (\O_h(x,y))=0\,. \ee Then the chiral physical vertex
operator is
%\be
%{\cal W}^{-1}_{h-1} (x,y) =   {\cal O}_h(x,y) \psi(x)
%\Phi_{h} (x) V_{h-1}(y)\,,
%\label{wvertex}
%\ee
\be
\O^{0}_h = e^{-\phi} \O_h(x,y)  \psi(x)
\label{opsi}
\ee
By requiring the bounds (\ref{hbound}) and (\ref{jbound})  to be
satisfied, we find that there are $k-1$ operators~$\O_h$, with $h=1,\frac32,\ldots,\frac{k}{2}$.
Finally, we verify that  in the OPE
\be
T_F(z) \O_h(x,y;w)\psi(x;w) &\sim& (z-w)^{-2} \left(  D_x^{+}  - 2x D^3_x + x^2 D_x^- \right)
\O_h(x,y;w) %\Phi_h(x) V_{h-1}(y) \
\nn \\
  &+& { O}\left(\frac{1}{z-w}\right)
\nn
\\
&\sim& { O}\left(\frac{1}{z-w}\right)
\label{brstforw}
\ee
all the
double poles cancel. Note that in flat space, this last condition
imposes on a vertex like $\xi \cdot \psi \, e^{i k \cdot X}$ the
polarization constraint $\xi \cdot k =0$. Here  the polarization is
already fixed in~(\ref{phipsi}) by the \slr\ symmetry, in a way
which is automatically BRST invariant.

For the computation of the three-point functions
we will need the representation of this vertex operator in the $0$ picture.
Acting on $\O^{0}_h$  with the picture-changing operator
$e^{\phi}T_F$
we get
\be
\tilde{\O}^{0}_h  &=& \left( J(x)   + \frac{2}{k} \psi(x) \psi_A D^A_x
+ \frac{2}{k} \psi(x)  \chi_a P^a_y
\right) \O_h(x,y)  \\
&=&   \left( (1- h)\hat{\jmath}(x)  + j(x)
+ \frac{2}{k} \psi(x)  \chi_a P^a_y
\right) \O_h(x,y)
\label{zppo}
\ee
%\be
%\tilde{\O}^{\psi}_h  &=& J(x) \O_h(x,y)  + \frac{2}{k} \psi(x) \psi_A D^A_x \Phi_h (x) V_{h-1}(y)
%+ \frac{2}{k} \psi(x) \Phi_{h}(x) \chi_a P^a_y V_{h-1}(y) \nn \\
%&=&   \left( (1- h_3)\hat{\jmath}(x)  + j(x) \right) \O_h(x,y)
%+ \frac{2}{k} \psi(x) \Phi_{h}(x) \chi_a P^a_y V_{h-1}(y)
%\ee
where all the terms are normal ordered and in the second line we have used eq.(\ref{jxsplit}) and the identity
\be
\psi(x)\psi_A D^A_x = -\frac{k}{2}h \hat{\jmath}(x)\,.
\ee

\end{enumerate}

\item $\hat{h}=0, \, \hat{\jmath}= 1 $ \\
In this case we have $H=h$ and $J=j-1,j,j+1$. The analysis for the
three cases is similar to the $\hat{h}=1,\,\hat{\jmath}=0$ cases.
The only physical chiral operators correspond to $H=h=j+1=J$. Using
the results of section
\ref{Tensor} on the tensor product of spin $j$ and spin
$1$ $SU(2)$ representations,  the physical chiral vertex is given by
\be
\O_h^{2}= e^{-\phi} \O_h(x,y) \chi(y)
\ee
where $\chi(y)$ is
the fermionic $SU(2)$ primary with $j=1$  defined in (\ref{chidef}).
The absence of double poles in the OPE of  $T_F$ with $\O_h(x,y)
\chi(y)$ is verified similarly to (\ref{brstforw}), and the number
of $\O^{2}_h$ operators is again $k-1$, as the number of
$\O_h(x,y)$  operators.

In the $0$ picture, the operator is
\be
\tilde{\O}_h^{2} &=&
\left( K(y)  + \frac{2}{k} \chi(y) \chi_a P^a_y   + \frac{2}{k} \chi(y)  \psi_A D^A_x
\right) \O_h(x,y)   \\
&=& \left( h \hat{k}(y)   + k(y)+ \frac{2}{k} \chi(y)  \psi_A D^A_x
\right) \O_h(x,y)
\label{zpco}
\ee
%\be
%\tilde{\O}_h^{\chi} =
%K(y)  \O_h(x,y)   + \frac{2}{k} \chi(y) \chi_a P^a_y V_{h-1}(y)  \Phi_h (x)
%+ \frac{2}{k} \chi(y) V_{h-1}(y) \psi_A D^A_x \Phi_{h}(x)
%\ee
where in the second line we have used (\ref{kxsplit}) and the identity
\be
\chi(y) \chi_a P_y^a = j\frac{k}{2}\hat{k}(y) \,,
\ee
with $j=h-1$.

\item $\hat{h}= \hat{\jmath}=1$
\\
For this case we have $H=h-1,h,h+1$ and $J=j-1,j,j+1$,
so there are nine sectors.
One can check that in all the cases, the mass shell condition
\be
\Delta = -\frac{h(h-1)}{k} + \frac{j(j+1)}{k}
 +1  + \Delta_T + N = \frac12
\ee
cannot be satisfied without violating the bound (\ref{hbound}) on $h$ or the
condition $\Delta_T \geq 0$. So there are no chiral physical operators in these sectors.

\end{enumerate}

\subsection{Ramond sector \label{Rsector}}

The mass shell condition is now, in the $-1/2$ picture,
\be
\Delta = \frac58  -\frac{h(h-1)}{k} + \frac{j(j+1)}{k} + \Delta_{T} + N  = \frac58\,,
\ee
and in the Ramond sector $\hat{\jmath}=\hat{h}=1/2$, so we have
$H=h \pm 1/2$ and $J=j\pm 1/2$. One can check that the mass shell condition
is satisfied without violating the bound (\ref{hbound}) on $h$, only by $H=h-1/2=j+1/2=J$.
As shown in section
\ref{Tensor}, in the $(x,y)$ basis,
the tensor product corresponding to this case
is given simply by
\be
\Phi_{h}(x)V_{h-1}(y)S(x,y)
\ee
where $S(x,y)$ is the field defined in (\ref{sdef}), and which can be realized in our background in
the two forms $S^{\pm}(x,y)$ defined in (\ref{spdef}) and
(\ref{smdef}). Including the spin field for the $T^4$ factor, our candidates
for the Ramond vertex operators are
\be
\O_h(x,y)S^{\pm}(x,y) e^{  \frac{i}{2}(\ve_4 \hat{H}_4 + \ve_5 \hat{H}_5)  }\,.
\ee
We should now check the BRST invariance of these
operators, which in the $-1/2$ picture implies the absence of
$(z-w)^{-3/2}$ singularities in their
OPE with $T_F$. Using the expressions (\ref{tfdec})-(\ref{tfb}) for
$T_F$, it is easy to check that the combination $\O_h(x,y)S^-(x,y)$
is BRST invariant, due to a precise cancelation between the
coefficients of $(z-w)^{-3/2}$ in its OPEs with $T^{\a}_F$ and
$T^{\beta}_F$. On the other hand, $\O_h(x,y)S^+(x,y)$ is not BRST
invariant, since its OPE with $T_F^{\beta}$  has no $(z-w)^{-3/2}$
singularities to cancel those arising in its  OPE with
$T_F^{\a}$.\footnote{One can check that the coefficient of the
$(z-w)^{-3/2}$ singularity in the OPE between $T_F^{\a}$ and
$\O_h(x,y)S^+(x,y)$ is zero only at $h=1/2$. This lies  at the
boundary in the allowed range (\ref{hbound}) for $h$, where a
discrete representation becomes a continuous one, but violates the
range (\ref{jbound}) for $j$, since $j=h-1=-1/2$.}

The GSO projection (\ref{gso}) imposes the further constraint $\ve_4 \ve_5 = -1$,
so the physical chiral operators in the R sector are finally
\be
\O_h^{a} = e^{-\frac{\phi}{2}} \O_h(x,y) s^{a}_-(x,y) \,, \qquad \qquad a=1,2
\ee
where
\be
s^{1}_{\pm}(x,y) &=& S^{\pm}(x,y) e^{+ \frac{i}{2}(\hat{H}_4- \hat{H}_5)} \,, \\
s^{2}_{\pm}(x,y) &=& S^{\pm}(x,y) e^{- \frac{i}{2}(\hat{H}_4- \hat{H}_5)} \,.
\ee
In order to compute the two-point functions of $\O_h^{a}$, we will need their expressions in the~$-3/2$ picture, which are
\be
\tilde{\O}_h^{a} &=& - \frac{\sqrt{k}}{(2h-1)} e^{-\frac{3\phi}{2}} \O_h(x,y) s_{+}^a(x,y)
\label{th}
\ee
This expression can be checked to be correct by acting on it with the picture raising operator $e^{\phi}T_F$, which
yields $\O_h^{a}$ (only the term $T^{\alpha}_F$ in (\ref{tfdec}) has a nontrivial action).

\vskip 1cm
In summary, all the chiral operators are obtained, in the canonical $-1/2, -1$ pictures,
by multiplying
the basic field $\O_h(x,y)$ defined in (\ref{odef}) by any
of the operators $ e^{-\phi} \psi(x)$, $e^{-\phi} \chi(y)$ or  $e^{-\frac{\phi}{2}} s^{a}_-(x,y)$.
The anti-holomorphic part of
the operators is fixed by multiplying also by an anti-holomorphic
field
$e^{-\bar{\phi}}\bar{\psi}(\bx)$, $e^{-\bar{\phi}} \bar{\chi}(\by)$ or  $e^{-\frac{\bar{\phi}}{2}} \bar{s}^{\bar{a}}_-(\bx,\by)$.

%Comparing the quantum numbers of the bulk and boundary operators, a clear dictionary emerges.
%For the holomorphic sector, the results can be
%summarized in the following table.
%There is a similar antiholomorphic table, and the possible ways to combine
%holomorphic with anti-holomorphic quantum numbers also match between bulk and boundary.
%In all the cases the relation $n = 2h-1$ holds.
%We have assumed that $Q_1=1$,  and note that in the boundary there is one more operator
%of each type, since $n$ takes the $Q_1 Q_5=k$ values $n=1\ldots k$, while $h$ takes the $k-1$ values
%$h=1, \frac32, \ldots, \frac{k}{2}$.

\subsection{Tensoring Bosonic and Fermionic Representations of $SL(2,R)$ and $SU(2)$ \label{Tensor}}

In this section we will work out the tensor products between the
bosonic and fermionic representations of \slr\ and $SU(2)$ that
appear in the chiral operators.

Let us first obtain the $h-1$ representation appearing in the tensor
product of a bosonic representation of \slr\ with quantum number
$h$, and the spin $1$ representation provided by the free
fermions~$\psi^A$. We work in a normalization for the modes
$\Phi_{h,m}$ such that \be J^3_0 \Phi_{h,m} &=& m \Phi_{h,m}
\,\,\,\,,
\\
J^{\pm}_0 \Phi_{h,m} &=& (m \mp (h-1) ) \Phi_{h,m\pm1} \,\,. \ee The
operators also depend on an antiholomorphic index $\bm$ which we
omit. We wish to determine the Clebsch-Gordon coefficients in the
expansion \be (\psi\Phi)_{h-1,m} = a_m \psi^3 \Phi_{h,m} + b_m
\psi^+ \Phi_{h,m-1} + c_m \psi^- \Phi_{h,m+1} \,. \label{tensor} \ee
Acting on this operator with both sides of
$J^{-}_0 = j^{-}_0 +
\hat{\jmath}^{-}_0$ we get
\be
&&
(m+h-2) \left(a_{m-1} \psi^3
\Phi_{h,m-1} + b_{m-1} \psi^+ \Phi_{h,m-2} + c_{m-1} \psi^-
\Phi_{h,m} \right) = \label{jminuseq}
\\
&&
\qquad \qquad\qquad\qquad
= (a_m(m+h-1) + 2b_m) \psi_3\Phi_{h,m-1} + b_m(m+h-2)\psi^+\Phi_{h,m-2}
\nn
\\
&& \qquad \qquad \qquad \qquad\qquad
\qquad \qquad \qquad \qquad \qquad \qquad
+ (a_m + c_m(m+h))\psi^-\Phi_{h,m}\,.
\nn
\ee
A second equation is obtained by acting on $(\psi\Phi)_{h-1,m-1}$ with both sides of
$J^{+}_0 = j^{+}_0 + \hat{\jmath}^{+}_0$,
\be
&& (m-h +1)
\left(a_{m} \psi^3 \Phi_{h,m} + b_{m} \psi^+ \Phi_{h,m-1} + c_{m} \psi^- \Phi_{h,m+1} \right) =
\label{jpluseq}
\\
&&
\qquad \qquad
= (a_{m-1}(m-h) - 2c_m) \psi_3\Phi_{h,m} + (b_{m-1}(m-1-h)-a_{m-1})\psi^+\Phi_{h,m-1} \nn \\
&& \qquad \qquad \qquad \qquad\qquad \qquad \qquad \qquad +\, c_{m-1} (m-h+1))\psi^-\Phi_{h,m+1}\,.
\nn
\ee
Equating the coefficients of both sides of (\ref{jminuseq}) and (\ref{jpluseq}), we
get six homogeneous equations for the six coefficients $a_m,b_m,c_m, a_{m-1},b_{m-1},c_{m-1}$.
Inserting the resulting values in (\ref{tensor}), gives, up to an overall rescaling
\be
(\psi\Phi)_{h-1,m} = 2 \psi^3 \Phi_{h,m} - \psi^+ \Phi_{h,m-1} - \psi^- \Phi_{h,m+1} \,.
\ee
This expression can be recast in the $x$ basis as
\be
(\psi\Phi)_{h-1}(x)&=& \sum_{m}(\psi \Phi)_{h-1,m}x^{-h+1-m} \\
&=& (-\psi^+ +2x\psi^3 x -x^2 \psi^-) \times \sum_{m}\Phi_{h,m}\, x^{-h-m}  \\
&=& \psi(x) \Phi_h(x)\,.
\ee
One can check that this result holds both for discrete representations,
where the sum runs over a semi-infinite range ($m=h,h+1\ldots$ or $m=-h,-h-1\ldots$),
and for continuous representations, where the sum runs over an
infinite range ($m=\a +\mathbb{Z}, \, \a\in [0,1)$).

The spin $j+1$ $SU(2)$ representation in the tensor product between
a bosonic representation of spin $j$ and the spin $1$ representation
provided by the fermions is $\chi^a$ is similarly obtained. By
exploiting the action of $K_0^{\pm}= k_0^{\pm}+ \hat{k}_0^{\pm}$, and
using the normalization (\ref{k3action})-(\ref{kpmaction}),
we get%\footnote{This expression corrects  eq.(A.6) of \cite{Kutasov:1998zh}.}
\be (\chi V)_{j+1,m}= -\chi^+ V_{j,m-1} + 2\chi^3 V_{j,m} +
\chi^-V_{j,m+1}  \,. \ee In the isospin $y$ basis, this becomes \be
(\chi V)_{j+1}(y) &=& \sum_{m=-j-1}^{j+1} (\chi V)_{j+1,m}\,y^{-m+j+1} \,, \\
&=&  (-\chi^{+} +2y\chi^3 + y^2 \chi^{-}) \times \sum_{m=-j}^{j}  V_{j,m} \, y^{-m+j} \,, \\
&=& \chi(y) V_j(y) \,.
\ee

\ni Finally, the representation with \slr\ and $SU(2)$ spins
$(h-1/2, j+1/2)$ in the tensor product of representations with spins
$(h, j)$ and $({\bf \frac12, \frac12 })$ is obtained by acting with
both $J_0^{\pm}= j_0^{\pm}+ \hat{\jmath}_0^{\pm}$ and $K_0^{\pm}=
k_0^{\pm}+ \hat{k}_0^{\pm}$, and is given by \be
 (S\Phi V )_{
(h-1/2,m+1/2)  \atop (j+1/2,n+1/2) }& =&
  |++\rangle \Phi_{h,m} V_{j,n} \\
&& + \, |+-\rangle \Phi_{h,m} V_{j,n+1}
+ |-+\rangle \Phi_{h,m+1} V_{j,n}
+ |--\rangle \Phi_{h,m+1} V_{j,n+1}
\nn
%&=& \sum_{\ve_1,\ve_2= \pm 1} |\ve_1,\ve_2 \rangle \Phi_{h,m+\frac{(1-\ve_1)}{2}} V_{j,n+\frac{(1-\ve_2)}{2}}
\ee
In the $(x,y)$ basis this becomes
\be
(S\Phi V )_{h-1/2, j+1/2}(x,y) &=& \sum_{m} \sum_{n=-j-1}^j (S\Phi V )_{
(h-1/2,m+1/2)  \atop (j+1/2,n+1/2) } x^{-m-h} y^{-n+j} \\
&=& \left( |++ \rangle + y |+- \rangle + x|-+ \rangle + xy |-- \rangle \right) \times
 \\
&& \qquad \qquad \qquad \times \sum_{m} \Phi_{h,m} x^{-m-h} \sum_{n=-j}^{j} V_{j,n} y^{-n+j} \nn \\
&=& S(x,y) \Phi_h(x) V_j(y)\,.
\ee
%\be
%(\psi \phi_h)_{h-1}(x,\bar{x}) &=& \left(x \psi^3  -\frac12 \psi^+  -\frac12 \psi^{-} x^2   \right) \left(\bar{x} \bar
%{\psi}^3  -\frac12 \bar{\psi}^+  -\frac12 \bar{\psi}^{-} \bar{x}^2   \right)  \phi_h(x,\bar{x})
%$\\(\psi \phi_h)_{h}(x,\bar{x}) &=& \left( -\psi^3 D_x^3 + \frac{\psi^+ D_x^- + \psi^-D_x^+}{2} \right) \phi_h(x,\bar{x})
% \\
%(\psi \phi_h)_{h+1}(x,\bar{x}) &=& \left( \psi^3 (D_x^-D_x^+-x \partial_x) +  \frac12 \psi^+ \partial_x(x \partial
%_x) + \frac12\psi^- D^+_x(2h+x \partial_x)
%\right) \phi_h(x,\bar{x})
%\ee

\section{Interactions in  Generalized $SU(2)$ WZW Models\label{secondap}}
In order to better understand the cancelation between the factors in the three-point functions
$C_S$ and~$C_H$ of the $SU(2)$ and \h3 WZW models at levels $k-2$ and $k+2$ respectively,
we will see in this appendix that these two quantities are solutions of functional equations
that are related by a sort of  ''Wick rotation''.

For convenience let us define ($b=1/\sqrt{k}$)
\be
\a_i & \equiv & b h_i \qquad
\a \equiv b h = \a_1 + \a_2 + \a_3 \,,
\\
a_i &\equiv& b j_i \qquad
a\equiv b j = a_1 + a_2 + a_3
\,,
\ee
and
\be
c_{H}(\a_1,\a_2,\a_3)& \equiv & C_{H}(h_1,h_2,h_3) \,, \\
c_{S}(a_1,a_2,a_3)& \equiv & C_{S}(j_1,j_2,j_3) \,.
\ee
The three-point function $c_{H}(\a_1,\a_2,\a_3)$
is determined by requiring it to be a solution of the functional equations
\be
\frac{c_{H}(\a_1+b/2,\a_2,\a_3)}{c_{H}(\a_1-b/2,\a_2,\a_3)}
\frac{c^-_H(\a_1)}{c^+_H(\a_1)} &=&
\frac{\ga^2(b(2\a_1-b))   \ga(b(\a-2\a_1 - b/2)) \ga(1- b\a+3b^2/2)}
{\ga(b(\a -2\a_3-b/2)) \ga(b(\a -2\a_2-b/2)) }
\,,
\label{feqforch}
\ee
\be
\frac{c_{H}(\a_1+\bi/2,\a_2,\a_3)}{c_{H}(\a_1-\bi/2,\a_2,\a_3)}
\frac{\tilde{c}^-_H(\a_1)}{\tilde{c}^+_H(\a_1)}
 &=&
\label{feqforch-2} \\
&& \!\!\!\!\!\!\!\!\!\!\!\!\!\!\!\!\!\!\!\!\!\!\!\!\!\!\!\!\!\!\!\!\!\!\!\!\!\!\!\!\!\!\!\!
\frac{\ga(\bi2\a_1-b^{-2})  \ga(2\bi \a_1 -1) \ga(\bi(\a-2\a_1 - \bi/2)) }
{\ga(\bi(\a -2\a_3 -\bi/2))  \ga(\bi(\a-2\a_2-\bi/2)) \ga(\bi(\a-\bi/2-b)) } \,,
\nn
\ee
where $\ga(x)=\Ga(x)/\Ga(1-x)$.
These equations where obtained in \cite{Teschner:1997ft}
by imposing crossing symmetry on a four point function, with  one of the fields corresponding
to the degenerate primaries $h=-1/2$ or $h=-k/2$ of the  \sl\ current algebra (see also \cite{ Giveon:2001up}).
The functions $c^{\pm}_H(\a_1), \tilde{c}^{\pm}_H(\a_1)$ are special structure constants that appear in the fusion
of  these degenerate fields with a generic primary
\be
\Phi_{-1/2} \Phi_h &=& c^+_H(\a) [\Phi_{h-1/2}] + c^-_H(\a) [\Phi_{h+1/2}] \,, \\
\Phi_{-k/2} \Phi_h &=& \tilde{c}^+_H(\a) [\Phi_{h-k/2}] + \tilde{c}^-_H(\a) [\Phi_{h+k/2}] + \tilde{c}^{\times}_H(\a)[\Phi_{1-h+k/2}]\,,
\ee
where $[\Phi_h]$ denotes the primary field and all its current algebra descendants.
The special structure constants can be obtained by a perturbative
calculation \cite{Giveon:2001up,Mukherjee:2006zv,Pakman:2006pr},
and are given by\footnote{To obtain the special structure constants self-consistently without any
perturbative computation, one should apply to \h3\ the
method used for Liouville theory in \cite{Pakman:2006hm}. A functional  equation that constrains the special
structure constants was obtained  in \cite{Teschner:1997ft}.}
\be
c^+_H(\a_1)&=& \tilde{c}^+_H(\a_1) =1 \\
c^-_H(\a_1)&=& \nu \frac{\ga(b(2\a_1 -b))}{\ga(2b\a_1)} \\
\tilde{c}^-_H(\a_1) &=& \tilde{\nu} \frac{\ga(\bi(2\a_1 -b))}{\ga(2\bi\a_1)}
\ee
Plunging these values into (\ref{feqforch}) and (\ref{feqforch-2}) yields
\be
\frac{c_{H}(\a_1+b,\a_2,\a_3)}{c_{H}(\a_1,\a_2,\a_3)} &=&
\frac{(\nu)^{-1} \ga(2b\a_1)  \ga(b(2 \a_1 +b )) \ga(b(\a-2\a_1 - b))
}{\ga(b(\a -2\a_3)) \ga(b(\a -2\a_2)) \ga(b(\a-b)) }\,,
\label{eqforch}
\\
\frac{c_{H}(\a_1+\bi,\a_2,\a_3)}{c_{H}(\a_1,\a_2,\a_3)} &=&
\frac{(\tilde{\nu})^{-1} \ga(2\bi\a_1)  \ga(\bi(2 \a_1 +\bi )) \ga(\bi(\a-2\a_1 - \bi)) }
{\ga(\bi(\a -2\a_3))  \ga(\bi(\a-2 \a_2)) \ga(\bi(\a-b)) }\,.
\label{eqforch-2}
\ee
The solution of these functional equations is, up to a multiplicative constant,
\be
c_H(\a_1,\a_2,\a_3) = \left[\nu b^{-2b^2}\right]^{-h+1} \frac{1}{\up(\a -b)}
\prod_{i=1}^{3} \frac{\up (2 \a_i  ) }{\up (\a - 2\a_i)} \,.
\ee
with $\tilde{\nu}=\nu^{b^2}b^{-4}$.
After rescaling the operators as %\cite{Teschner:1999ug}
\be
\Phi_h \rightarrow \frac{\Phi_h}{\ga(b^2(2h-1))}
\label{rescaling}
\ee
we get the three-point function given in (\ref{ch}), up to a multiplicative constant.
Note that $c_H$ is not an analytic function of $b$,
since $\up$ has a branch cut for positive imaginary values of $b$ \cite{Zamolodchikov:2005fy}.

The conformal bootstrap method used
for the  \h3\ WZW model can also be applied in the  $SU(2)$
case.\footnote{The technique  used in \cite{Zamolodchikov:1986bd} to compute the $SU(2)$ WZW three-point functions
(\ref{zf3pf}) was  different, and consisted in  exploiting the
relation between generic  $SU(2)$ four-point function and a
five-point function of the minimal models.} Indeed,
the steps in \cite{Teschner:1997ft} that lead to the level $k+2$ \h3\ functional equations,
have level $k-2$ $SU(2)$ counterparts  which are simply obtained  by the replacements
\be
&h& \rightarrow  -j \,,
 \label{repl1}
\\
&b& \rightarrow -ib \,,
\label{repl2}
\\
&\bi& \rightarrow i \bi \,.
\label{repl3}
\ee
Applying this to  (\ref{feqforch}) and (\ref{feqforch-2})
 we get the functional equations
\be
\frac{c_{S}(a_1-b/2,a_2,a_3)}{c_{S}(a_1+b/2,a_2,a_3)}
\frac{c^-_S(a_1)}{c^+_S(a_1)} &=&
\frac{\ga^2(b(2a_1+b))   \ga(b(a-2a_1 + b/2))}
{\ga(b(a -2a_3+b/2)) \ga(b(a -2a_2+b/2)) \ga(b(a+3b/2))} \,\,\,
\label{feqfors}
\ee
\be
\frac{c_{S}(a_1+\bi/2,a_2,a_3)}{c_{S}(a_1-\bi/2,a_2,a_3)}
\frac{\tilde{c}^-_S(a_1)}{\tilde{c}^+_S(a_1)}
 &=&
\label{feqfors-2} \\
&& \!\!\!\!\!\!\!\!\!\!\!\!\!\!\!\!\!\!\!\!\!\!\!\!\!\!\!\!\!\!\!\!\!\!\!\!\!\!\!\!\!\!\!\!
\frac{\ga(\bi(-2a_1+\bi))  \ga(-2\bi a_1 -1) \ga(\bi(-a+2a_1 + \bi/2)) }
{\ga(\bi(-a +2a_3 +\bi/2))  \ga(\bi(-a+2a_2+\bi/2)) \ga(\bi(-a+\bi/2-b)) } \,.
\nn
\ee
Under (\ref{repl1})-(\ref{repl3}), the degenerate primaries go to $j=1/2$ and $j=-k/2$.
The latter does not belong to the standard spectrum of the $SU(2)$ WZW model, but
it is  a degenerate vector of the $SU(2)$ affine algebra \cite{Kac:1979fz}.
The fusion rules are now \cite{Awata:1992sm}
\be
V_{1/2} V_j  &=& c^+_S(a) [V_{j+1/2}] + c^-_S(a) [V_{j-1/2}] \,, \\
V_{-k/2} V_j &=& \tilde{c}^+_S(a) [V_{j-k/2}] + \tilde{c}^-_S(a) [V_{j+k/2}] + \tilde{c}^{\times}_S(a)[V_{-1-j+k/2}]\,.
\ee
Since in the case of the $SU(2)$ WZW model, the primaries are normalized as
\be
\langle V_{j_1}(y_1) V_{j_2}(y_2) \rangle = \delta_{j_1,j_2}|y_{12}|^{2j_1},
\ee
we can identify the special structure constants with particular three-point functions,
\be
c^{\pm}_S(a_1)&=& c_S(a_1,b/2,a_1\pm b/2) \,,\\
\tilde{c}^{\pm}_S(a_1) &=& c_S(a_1,-\bi/2,a_1 \mp \bi/2) \,.
\label{spsc}
\ee
This in turn allows to determine them from (\ref{feqfors}) and (\ref{feqfors-2}).
Specializing (\ref{feqfors}) to $a_3=a_1, a_2=b/2$,  and (\ref{feqfors-2}) to $a_3=a_1, a_2=-b^{-1}/2$  we get
\be
\left( \frac{c^-_S(a_1)}{c^+_S(a_1)} \right)^2 = \frac{\ga^2(b(2a_1+b))}{\ga(2ba_1) \ga(2b(a_1+b)) }
\ee
and
\be
\left( \frac{\tilde{c}^-_S(a_1)}{\tilde{c}^+_S(a_1)} \right)^2
= \frac{\ga(-2a_1\bi + b^{-2}) \ga(-2 a_1 \bi -1)}{\ga(-2 a_1 \bi) \ga(\bi(-2a_1+\bi-b)) } \,.
\ee
Plunging these values back into (\ref{feqfors}) and (\ref{feqfors-2}) gives the
functional equations
\be \frac{c_{S}(a_1+b, a_2, a_3)}{c_{S}(a_1,a_2, a_3)} =
\frac{  \ga(b(a-2 a_2+b))\ga(b(a-2 a_3+b))\ga(b(a + 2b))}
{ \ga(b(2a_1+2b))[\ga(b(2a_1+b))\ga(b(2a_1+3b))]^{1/2}\ga(b(a -2 a_1))}\,,
\label{eqforcs}
\ee
\be
\frac{c_{S}(a_1+\bi, a_2,a_3)}{c_{S}(a_1, a_2, a_3)}
&=& \frac{  \ga(-\bi(a -2 a_1-\bi))}
{\ga(-\bi(a - 2 a_2))\ga(-\bi(a-2 a_3))\ga(-\bi(a+b))} \times
\label{eqforcs-2} \\
&& \frac{1}{[\ga(2+ 2 a_1 \bi) \ga(1+ 2a_1\bi + b^{-2}) \ga(1+2a_1\bi) \ga(2+2a_1\bi+b^{-2})]^{1/2}}
\nn
\,.
\ee
Using (\ref{shiftu}), it follows that the above two functional equations %(\ref{eqforcs}) and (\ref{eqforcs-2})
are solved by
\be
c_S(a_1,a_2,a_3) =  \frac{\sqrt{\ga(b^2)} b^{\frac12 -b^2}}{\up(b)} \up (a + 2b) \prod_{i=1}^{3} \frac{\up (a-2a_i +b ) }
{[\up (2a_i +b) \up (2a_i +2b)]^{1/2}} \,,
\label{su3pf}
\ee
where we have fixed the arbitrary constant by requiring $c_S(a_1,a_1,0)=1$.
Using (\ref{psu})  one can express  $c_{S}(a_1, a_2, a_3)$ in terms of $P(s)$, and it is
immediate  to check that the resulting  expression for (\ref{su3pf})  precisely
coincides with $C_{S}(j_1, j_2, j_3)$ in (\ref{zf3pf}) - up to the factor $N_{j_1,j_2,j_3}$ which
should be added to~(\ref{su3pf}).

The above form of the $SU(2)$ three-point functions is
defined for any value of~$2j_i$, not only for positive
integers.\footnote{Indeed, we have used this freedom in (\ref{spsc}), since we have
evaluated $c_S$ at a value corresponding to $j_2=-k/2 < 0$.} Thus $c_S(a_1,a_2,a_3)$ are the
three-point functions  of a {\it generalized} $SU(2)$ WZW model,
similar to the generalized minimal models studied
in~\cite{Zamolodchikov:2005fy,Kostov:2005av}.\footnote{V.~Petkova has pointed out to us that
it should be possible to obtain the structure constants for these generalized $SU(2)$ from
those of the generalized minimal models by using the relation between the structure constants in
both models found in \cite{Furlan:1996vu}. }
In those works, similar relations  and cancelations between Liouville theory
and the minimal models three-point functions were observed.

Finally, note that to obtain  $c(\a_1,\a_2,\a_3) \equiv  c_H(\a_1,\a_2,\a_3)c_S(\a_1-b,\a_2-b,\a_3-b)$ in (\ref{3pp}),
instead of multiplying the two expressions,
we could have  combined eqs. (\ref{eqforch}), (\ref{eqforch-2}), (\ref{eqforcs}) and (\ref{eqforcs-2}).
This gives the functional equations,
\be
\frac{c(\a_1+b,\a_2,\a_3)}{c(\a_1,\a_2,\a_3)} &=&
(\nu)^{-1}
\left[ \frac{  \ga(b(2\a_1+b))}{\ga(b(2\a_1-b))} \right]^{1/2}\,,
\\
\frac{c(\a_1+\bi,\a_2,\a_3)}{c(\a_1,\a_2,\a_3)}
&=&
(\tilde{\nu} )^{-1}
\left[ (2\a_1 \bi -1)^2
(2\a_1\bi+ b^{-2}-1)^2
\right]^{1/2} \,,
\ee
whose solution, after the rescaling (\ref{rescaling}),
is given by (\ref{3pp}), up to a constant.

\newpage

\bibliography{h3bib}

\bibliographystyle{JHEP}
%\bibliographystyle{unsrt}
%\bibliography{ref}
%\bibliographystyle{apsrev}
%\bibliographystyle{plain}
%\bibliographystyle{utphys}

\end{document}